\documentclass[twocolumn,twoside]{IEEEtran}

\ifCLASSINFOpdf

\else

\fi

\ifCLASSOPTIONcompsoc
    \usepackage[caption=false, font=normalsize, labelfont=sf, textfont=sf]{subfig}
\else
    \usepackage[caption=false, font=normalsize]{subfig}
\fi
\usepackage{lipsum}%
\usepackage[dvipsnames]{xcolor}
\usepackage{algorithm,algorithmic}
\usepackage{balance}
\usepackage{multicol}   % for equalize of last page columns
\usepackage{cite}
\usepackage{gensymb}
\usepackage{multirow}
\usepackage{graphics}
\usepackage{epsfig}
\usepackage{graphicx}
\usepackage{epstopdf}
\usepackage{textcomp}
\usepackage{amsmath}
\usepackage{mathtools}
\usepackage{filecontents}
\usepackage{lipsum,color}
\usepackage{amssymb}
\usepackage{float}
\usepackage{colortbl} % for painting tables
\usepackage{times} % assumes new font selection scheme installed
\usepackage{amsthm}  % for proof
\usepackage{amsfonts}

\theoremstyle{break}
\begin{document}
\title{3D Channel Characterization and Performance Analysis of UAV-Assisted Millimeter Wave Links   }

%------------ Junil Choi student of Heath
%------------ Kaibin Huang with Androws
%------------ Rui Zhang

%Xiang Cheng
\author{Mohammad~Taghi~Dabiri,~Mohsen~Rezaee,~Behrouz~Maham,~{\it Senior Member,~IEEE}, \\
	~Walid~Saad,~{\it Fellow,~IEEE},~and~Choong~Seon~Hong,~{\it Senior Member,~IEEE}
\thanks{Mohammad Taghi Dabiri and Mohsen Rezaee are with the ICT research institute, Iran Telecommunication Research Center (ITRC), Tehran, Iran (E-mail: m.dabiri@itrc.ac.ir, m.rezaeekh@itrc.ac.ir).}
 	
%\thanks{Mohsen Rezaee is with the ICT research institute, Iran Telecommunication Research Center (ITRC), Tehran, Iran (E-mail: m.rezaeekh@itrc.ac.ir).}	 
	
\thanks{Behrouz Maham is with the Department of Electrical and Computer Engineering, School of Engineering and Digital Sciences, Nazarbayev University, Astana 010000 , Kazakhstan  (Email: behrouz.maham@nu.edu.kz). }
	
\thanks{Walid Saad  is with the Bradley Department of Electrical and Computer Engineering, Virginia Tech, USA (E-mail: walids@vt.edu).}
	
\thanks{Choong Seon Hong is with Department of Computer Science and Engineering, Kyung Hee University, South Korea (E-mail: cshong@khu.ac.kr). }
}

\maketitle

%%%%%%%%%%%%%%%%%%%%%%%%%%%%%%%%%%%%%%%%%%%%%%%%%%%%%%%%%%
%%%%%%%%%%%%%%%%%%%%%%%%%%%%%%%%%%%%%%%%%%%%%%%%%%%%%%%%%%
\begin{abstract}
%%%%%%%%%%%%%%%%%%%%%%%%%%%%%%%%%%%%%%%%%%%%%%%%%%%%%%%%%%
%%%%%%%%%%%%%%%%%%%%%%%%%%%%%%%%%%%%%%%%%%%%%%%%%%%%%%%%%%
Recently, the use of millimeter wave (mmW) frequencies has emerged as a promising solution for wirelessly connecting unmanned aerial vehicles (UAVs) to ground users.
%the fronthaul links of wireless networks that rely on unmanned aerial vehicles (UAVs) acting as access points.
However, employing UAV-assisted directional mmW links is challenging due to the random fluctuations of hovering UAVs.
%+is becoming crucial. 
% can play a key role in enabling
%
%The main contribution of this paper is to analyze UAV-based mmW links when UAVs are equipped with square array antenna.
In this paper, the performance of UAV-based mmW links is investigated when UAVs are equipped with square array antennas.
The 3GPP antenna propagation patterns are used to model the square array antenna.
It is shown  that the square array antenna is sensitive to both horizontal and vertical angular vibrations of UAVs.
In order to explore the relationship between the vibrations of UAVs and their antenna pattern, 
the UAV-based mmW channels are characterized by considering the large scale path loss, small scale fading along with antenna patterns as well as the random effect of UAVs' angular vibrations.
To enable effective performance analysis, tractable and closed-form statistical channel models are derived for aerial-to-aerial (A2A), ground-to-aerial (G2A), and aerial-to-ground (A2G) channels. The accuracy of analytical models is verified by employing Monte Carlo simulations. 
Analytical results are then used to study the effect of antenna pattern gain under different conditions for the UAVs' angular vibrations for establishing reliable UAV-assisted  mmW links in terms of achieving minimum outage probability.
Simulation results show that the performance of UAV-based mmW links with directional antennas is largely dependent on the random fluctuations of hovering UAVs.
Moreover, UAVs with higher antenna directivity gains achieve better performance at larger link length. However, for UAVs with lower stability, lower antenna directivity gains result in a more reliable communication link. 
The developed results can therefore be applied as a benchmark for finding the optimal antenna directivity gain of UAVs under the different levels of instability without resorting to time-consuming simulations.

%
%Finally, as a practical scenario, the optimal design of UAV-assisted mmW relay link is studied.
% when the ground source and destination nodes are blocked by obstacles.
%Based on the geometrical properties of a given region such as height of obstacles and their positions with respect to the source and destination, the optimal antenna pattern along with the optimal aerial position that allows a UAV relay to attain minimum outage probability are investigated.
\end{abstract}
\IEEEpeerreviewmaketitle

%%%%%%%%%%%%%%%%%%%%%%%%%%%%%%%%%%%%%%%%%%%%%%%%%%%%%%%%%%%%
%%%%%%%%%%%%%%%%%%%%%%%%%%%%%%%%%%%%%%%%%%%%%%%%%%%%%%%%%%%%
\section{Introduction}
%%%%%%%%%%%%%%%%%%%%%%%%%%%%%%%%%%%%%%%%%%%%%%%%%%%%%%%%%%%%
%%%%%%%%%%%%%%%%%%%%%%%%%%%%%%%%%%%%%%%%%%%%%%%%%%%%%%%%%%%% https://www.hindawi.com/journals/wcmc/2017/1830987/
\IEEEPARstart{N}{ext}-generation
cellular networks will inevitably rely on two technologies, high-frequency millimeter wave (mmW) bands and unmaned aerial vehicles (UAVs) \cite{saad2019vision,chen2017caching,mozaffari2017mobile,mozaffari2018tutorial,khawaja2019survey,cao2018airborne,mozaffari2018beyond,amer2019mobility}.
%
%Although mmW bands can provide a large bandwidth for wireless transmission, mmW signals suffer from blockage. As such, in order to fully exploit the benefit of mmW frequencies, the receiver (Rx) must be placed within the line-of-sight (LoS) of the transmitter (Tx) \cite{mezzavilla2018end}. Due to the impracticality of establishing long LoS links, specially in dense urban environments, the terrestrial mmW links are mostly limited to short-range links. Compared to the terrestrial mmW link, the advantage of using UAV-based mmW communications is the ability to adjust the position in the sky with respect to the obstacles, and enhance the reliability by offering the LoS component between Tx and Rx \cite{mozaffari2018beyond}.
%
The advancement of UAV  technologies and their reducing cost, have made future cellular systems more likely to be equipped with UAVs as flying base stations (BSs)  which effectively enhances the network flexibility and capacity \cite{zhang2018cellular}.
%Meanwhile, ultra dense small cell networks face many challenges in terms of backhaul, interference, and overall network modeling.
Meanwhile, flying BSs mounted on UAVs require high capacity backhaul links \cite{fouda2019interference}.
%
%For tethered UAVs, optical fiber backhauling is an approach for connecting aerial BS to a core network. 
%Unlike high capacity, employing optical fiber can be expensive and limited the maneuverability  
%especially as backhaul of aerial BSs or mobile relay stations,
More recently, mmW backhauling has been proposed as a promising approach for connecting aerial BSs to a core network because of three reasons.
First, unlike ground nodes that suffer from blockage, the flying nature of aerial nodes offers a higher probability of line-of-sight (LoS) between communication nodes.
% unlike the scarce available bandwidth at microwave frequency bands,
Second, the large available bandwidth at mmW bands can provide high capacity point-to-point aerial communication links as needed for the backhaul of aerial BSs.
Third, the small wavelength enables the realization of a compact form of beam-steerable, highly directive antenna arrays on a small UAV with limited payload to compensate for the high path-loss of the mmW band \cite{yu2018capacity}. 

To exploit the advantages of UAV equipped with directional mmW antennas, it is important to have a comprehensive and accurate channel model while taking into account the antenna propagation pattern  as well as the random effect of a UAV's angular vibrations.
%%%%%%%%%%%%%%%%%%%%%%%%%
%++++Enabling advanced multiantenna and multiplexing schemes with efficient use of available spectrum has been a subject of extensive research. 
Although channel modeling in the context of UAV communications has been studied in recent works \cite{khuwaja2018survey,yan2019comprehensive,matolak2017air,sun2017air}, these studies are limited to sub-6 GHz bands which cannot be directly extended to mmW systems.
%%%%%%%%%%%%%%%%%%%%%%%%%
%%%%%%%%%%%%%%%%%%%%%%%%%
Meanwhile, most of the prior studies on mmW communications \cite{semiari2017caching,rappaport2013millimeter,wu201760} do not address the presence of UAVs, with the exception of a few recent works in \cite{khawaja2017uav,khawaja2018temporal,xiao2016enabling,kovalchukov2018analyzing,rupasinghe2019non,rupasinghe2019angle,galkin2018backhaul,gapeyenko2018flexible}.
%
%The channel modeling of UAV-based mmW communications have been the subject of several recent works.
%
%For instance, in \cite{khawaja2017uav}, the authors propose a ray tracing approach to characterize the mmW propagation channel for an air-to-ground link  at  28 GHz and 60 GHz. Meanwhile, the work in \cite{khawaja2018temporal} studies the small scale temporal and spatial characteristics of mmW air-to-ground LoS propagation channels at 28 GHz under different conditions.
For instance, the works in \cite{khawaja2017uav} and \cite{khawaja2018temporal} study a ray tracing approach to characterize the mmW propagation channel for an air-to-ground link  at  28 GHz and 60 GHz under different conditions.
%A single, omni-directional antenna
%%%%%%%%%%%%%%%%%%%%%%%%%
%%%%%%%%%%%%%%%%%%%%%%%%%
%However, the authors in \cite{khawaja2018temporal} assumed half-wave dipole antennas with an omni-directional pattern and, hence, their approach does not properly capture the UAV fluctuation effects.
However, the results of these works are obtained with the assumption of half-wave dipole antennas with an omni-directional pattern.
%%%%%%%%%%%%%%%%%%%%%%%%%
%+due to its sensitivity to the atmosphere and severe propagation loss, which is inversely proportional to the squared wavelength, thereby affecting long range transmission [7].
The reliability and performance of UAV-based mmW links can be severely affected by impairments such as sensitivity to the atmospheric conditions and large propagation loss.
%, which is inversely proportional to the squared wavelength.
%can establish high directional gain that are needed to cope with sever propagation loss of long link length.
Due to the UAVs' transmission power constraints, using antennas with high gain is needed to combat severe propagation loss, particularly for longer links\footnote{
%With the small wavelength and recent advances in modular antenna array technology for high frequency bands, high beamforming gains can be obtained to combat large propagation loss and therefore are potentially suitable for 
%
Advances in fabrication of antenna array technology at mmW bands allow the creation of large antenna arrays with high gain in a cost effective and compact form. For instance, , light-weight directional mmW array antennas (e.g., less than 1 kg) are already available in the market, which are suitable to be mounted on UAVs with limited payload.}.

Directional UAV-based mmW communications have been the subject of more recent works such as \cite{xiao2016enabling} in which the authors study
the directional mmW channel characteristics for UAV networks  by considering the Doppler effect as a result of UAV movement.
%In \cite{kovalchukov2018analyzing}, the authors introduce a novel model for UAV-based mmW communication composes the high directionality of transmissions along with the UAV's random heights.
The effects of directionality and random heights in UAV-based mmW communications are studied in \cite{kovalchukov2018analyzing}.
In \cite{rupasinghe2019non} and \cite{rupasinghe2019angle}, the authors propose an analytical framework for non-orthogonal multiple-access transmission with UAVs so as to support more users in a hotspot area such as a football stadium.
%%%%%%%%%%%%%%%%%%%%%%%%%%
%\textcolor{blue}{ However, the proposed channel models in \cite{xiao2016enabling,kovalchukov2018analyzing,rupasinghe2019non,rupasinghe2019angle} are only applicable for a link between users and a perfectly stable UAV.}
%The directional UAV-based mmW link can be also used as a high powerful wireless backhaul link between aerial UAV-based access point and ground core network \cite{zhang2019research}.
In \cite{galkin2018backhaul}, the authors use stochastic geometry to study directional UAV-based backhaul links operating at 2 GHz and 73 GHz. For simplicity, in \cite{galkin2018backhaul}, the antenna pattern is approximated by a rectangular radiation pattern.
%The use of aerial relay nodes is investigated in \cite{gapeyenko2018flexible} to propose a flexible and re-configurable UAV-assisted backhaul operation that takes into account the dynamic blockage of mmW links and the heterogeneous mobility of signal blockers.

In \cite{gapeyenko2018flexible}, the authors study a UAV-based communication system that takes into account the dynamic blockage of mmW links.
%In \cite{han2018connectivity} the authors proposed an analytical framework to define and characterize the connectivity in a mmW air-to-everything links as a function of the altitude of UAV and different parameters of buildings including their densities, sizes, and heights.
%%%%%%%%%%%%%%%%%%%%%%%%%%%
%should be distinctively characterized in terms of a UAV's random vibrations due to hovering fluctuations and also mmW propagation characteristics.
%
%random vibrations of UAVs can lead to antenna gain mismatch between transmitter and receiver which, in turn, can cause signal-to-noise ratio (SNR) fluctuations at the receiver side that can significantly degrade the reliability of the system. 
High directional mmW communication systems suffer from misalignment between transmitter (Tx) and receiver (Rx). Due to the payload limitations for employing high quality antenna stabilizers, careful alignment is not practically feasible in aerial links, particularly, for small multi-rotor UAVs. This leads to an unreliable communication system due to antenna gain mismatch between transceivers \cite{guan2019effects,zhong2019adaptive,pokorny2018concept}.
%
%
%However, the works in \cite{galkin2018backhaul,gapeyenko2018flexible} ignore the random fluctuations of UAVs and assume a perfect beam alignment between transceivers.
However, the results of these works are obtained by neglecting the effect of UAVs' random fluctuations.
More recently, the authors in \cite{dabiri2019analytical}, studied the problem of channel modeling for directional UAV-based mmW links including the effects of angle-of-arrival (AoA) and angle-of-departure (AoD) fluctuations. The results of \cite{dabiri2019analytical} clearly demonstrate that the orientation fluctuations of hovering UAVs degrade the performance of directional UAV-based mmW links, significantly. However, the results of \cite{dabiri2019analytical} are obtained for a simplified state of uniform linear array (ULA) of antennas.
Although ULA antennas have lower complexity, the square array antenna outperforms ULA antennas.
%The square array antenna outperform than ULA antennas with expense of more complexity.  

%-----------------------------------------------
%-----------------------------------------------
\subsection{Major Contributions}
%-----------------------------------------------
%-----------------------------------------------
The main contribution of this paper is the derivation of novel analytical channel models for UAV-based mmW links and performance analysis of UAV-based mmWave communication systems when UAVs are equipped with square array antennas.
In particular, we consider balloon and rotary-wing UAVs such as quadrotor drones that can hover and remain stationary over a given area in the sky.
In addition to the large scale path loss and small scale fading, we show the channel of UAV-based mmW links will be a function of UAV's angular fluctuations along with the shape of antenna patterns.
%+++are remarkably diverse
Unlike the ULA antenna that is resistant to a UAV's horizontal fluctuations, we will show that the square array antenna is sensitive to both horizontal and vertical fluctuations, and thus, the channel characterization of UAVs equipped with square array antennas is remarkably different from ULA antenna case. 
%Due to the sensitivity of directional mmW link performance to both horizontal and vertical antenna vibrations, the channel characterization of UAVs equipped with square array antennas is remarkably different from ULA antenna case. 
%both horizontal and vertical fluctuations must be taken into account.
In summary, our key contributions include: 
\begin{itemize}
	\item We characterize the precise channel models for three UAV-based mmW communication links: aerial-to-aerial link (called A2A link),  ground-to-aerial link (called G2A link), and  aerial-to-ground link (called A2G link).
	By taking into account the 3GPP antenna propagation patterns, the actual channel models are characterized in presence of large scale path loss and small scale fading along with the influence of UAV angular fluctuations.
	%%%%%%%%%%%%%%%%%%%%%%%
	\item Then, for the characterized channels, we derive closed-form analytical expressions for A2A, A2G, and G2A channels. 
	%
	%In addition to the tractability, the characteristics of UAV-based mmW links such as antenna pattern shape and severity of UAVs' vibrations, can be easily set into the provided analytical channel models.
	%
	Then, by providing Monte Carlo simulations for actual UAV-based mmW channels, the accuracy of the derived analytical expressions is verified.
	%%%%%%%%%%%%%%%%%%%%%%%
	\item We also derive the closed-form expressions for the outage probability of the considered UAV-based mmW links. The accuracy of the analytical expressions is verified by using simulations.
	%
	%For any given severity of UAVs' vibrations, optimizing the radiation pattern requires balancing an inherent tradeoff between increasing directional gain to compensate for the large path loss at mmW frequencies and decreasing it to alleviate the adverse effect of transceivers vibrations.
	For any given strength of UAVs' vibrations, optimizing radiation pattern shape requires balancing an inherent tradeoff between
	decreasing pattern gain to alleviate the adverse effect of a UAV's vibrations and increasing it to compensate the large path loss at mmW frequencies. 
	The analytical results are applied as a benchmark for finding optimal antenna pattern shapes mounted on UAVs under the different levels of UAVs' instability without resorting to time-consuming simulations.
	%%%%%%%%%%%%%%%%%%%%%%%
	%\item Then, as a practical scenario, the optimal design of UAV-assisted mmW relay link is investigated when the ground source and destination nodes are blocked by obstacles. Based on the geometrical properties of a given region such as height of obstacles and their positions with respect to the source and destination, by using time consuming Monte Carlo simulations, we find the optimal directivity antenna gain along with the best aerial position for UAV relay to attain minimum outage probability. Then, as a suboptimal method, the optimization parameters are obtained by using the provided analytical method. The obtained results show that the analytical suboptimal method achieves a performance that is close to the optimal method while reduces the processing time, significantly.
\end{itemize}

The rest of this paper is organized as follows. In Section II, we characterize the actual channel models between UAVs.
Then, in Section III, we provide the analytical channel models.
Next, in Section IV, we provide the simulation results to verify the derived analytical channel models and study the link performance and antenna pattern optimization. 
Finally, conclusions are drawn in Section V.

\section{System Model}
%%%%%%%%%%%%%%%%%%%%%%%%%%%%%%%%%%%%%%%%%%%%%%%%%%%%%%%%%%%%%%%%%%%%%%%%%%%
%%%%%%%%%%%%%%%%%%%%%%%%%%%%%%%%%%%%%%%%%%%%%%%%%%%%%%%%%%%%%%%%%%%%%%%%%%%

%In this section, we introduce the system model. %Our notation is summarized in Table \ref{tabs}.

\begin{figure}
	\begin{center}
		\includegraphics[width=3.4 in ]{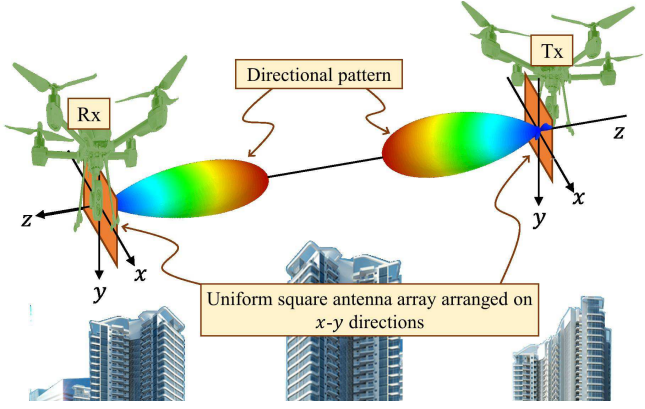}
		\caption{Illustration of an A2A link. The $z$ axis refers to the direction that extends from the Tx to the Rx node. The hovering UAVs adjust their antenna main lobes in the direction of $z$-axis.}
		\label{st1}
	\end{center}
\end{figure}

\begin{figure}
	\begin{center}
		\includegraphics[width=3.2 in ]{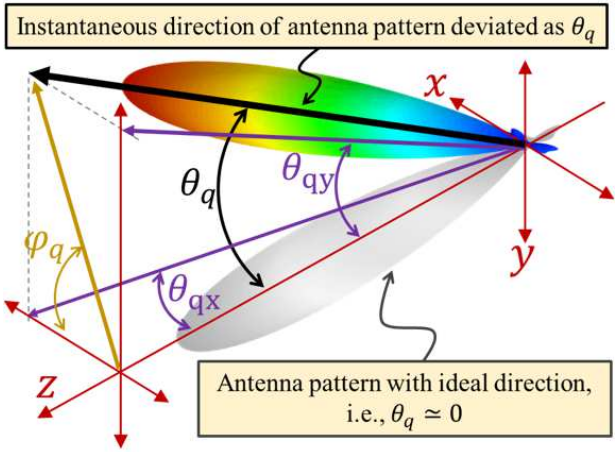}
		\caption{A graphical example of UAV orientation fluctuations. Here, the direction of antenna's main-lobe deviates by $\theta_q$ from the $z$ axis. As illustrated, $\theta_{qx}$ and $\theta_{qy}$  are the instantaneous orientation fluctuations in $x-z$ and $y-z$ Cartesian coordinates, respectively.}
		\label{st2}
	\end{center}
\end{figure}

We consider a UAV-based mmW communication link that is used to provide a high capacity point-to-point link.
This link can be A2A, A2G, or G2A.
In an A2A link, the Tx and Rx antennas are mounted on two hovering UAVs, in an A2G link, the Tx and Rx antennas are mounted on a UAV and a ground station, respectively, and in a G2A link, the Tx and Rx antennas are mounted on a ground station and a UAV, respectively.  
We use the subscript $q\in\{t,r\}$ to denote respectively, the Tx and Rx antenna. For instance, let $\sigma^2_{qx}$ and $\sigma^2_{qy}$ be the standard deviations of UAV orientation fluctuations in the $x-z$ and $y-z$ axes, respectively. Thus, $\sigma^2_{tx}$ and $\sigma^2_{ty}$ represent the standard deviations of the Tx in the $x-z$ and $y-z$ axes, respectively, and $\sigma^2_{rx}$ and $\sigma^2_{ry}$ represent the standard deviations of the Rx in the $x-z$ and $y-z$ axes, respectively.
In our point-to-point communication link, the UAVs are hovering in space at a distance of $Z$ from each other. The position of Tx and Rx  are respectively located at $[0, 0, 0]$ and $[0, 0, Z]$ in Cartesian coordinate system $[x,y,z]\in \mathcal{R}^{1\times3}$ and are known at the transceiver.%, which can be shared by transmitting  periodic broadcast messages.
As shown in Fig. \ref{st1}, $z$ axis refers to the direction that extends from Tx toward Rx node. The hovering UAV sets the main-lobe of array antenna pattern in the direction of $z$-axis as depicted in Fig. \ref{st1}. In practice, the instantaneous orientation of a UAV can randomly deviate from its means denoted by $\theta_q$. This, in turn, leads to deviations in the AoD of Tx and/or AoA of Rx antenna pattern.
As shown in Fig. \ref{st2}, the antenna orientation fluctuations are denoted by $\theta_{qx}$ and $\theta_{qy}$ in the $x-z$ and $y-z$ Cartesian coordinates, respectively. In particular, at the  Tx side, the AoD deviations are denoted by $\theta_{tx}$ and $\theta_{ty}$ in $x-z$ and $y-z$ Cartesian coordinates, respectively, and at the Rx side, the AoA deviations are denoted by $\theta_{rx}$ and $\theta_{ry}$ in $x-z$ and $y-z$ Cartesian coordinates, respectively. The UAV-based mmW links are grouped into three categories: a) A2A link between two UAVs, b) G2A link between a ground Tx and a UAV Rx, and c) A2G link between a UAV Tx and a ground Rx. In practice, the ground nodes have negligible orientation fluctuations and, hence, , we assume $\theta_q\simeq0$.
Based on the central limit theorem, the deviations of the UAVs' orientations are considered to be Gaussian distributed \cite{dabiri2018channel,kaadan2014multielement,dabiri2019tractable}. 
%
%From the literature, we assume the variance of orientation fluctuations of the UAV node is approximately same in $x-z$ and $y-z$ planes, i.e., $\sigma^2_\textrm{qx}\simeq\sigma^2_\textrm{qy}=\sigma^2_\textrm{qo}$. 
Therefore, we have $\theta_{qx}\sim \mathcal{N}(\theta'_{qx},\sigma^2_{qo})$, and $\theta_{qy}\sim \mathcal{N}(\theta'_{qy},\sigma^2_{qo})$, where $\theta'_{qx}$ and $\theta'_{qy}$ are the boresight direction of the antennas in $x-z$ and $y-z$ Cartesian coordinates, respectively, and $\sigma^2_{qx}\simeq\sigma^2_{qy}=\sigma^2_{qo}$.

%++++++\textcolor{red}{The antennas are tilted up towards the sky, to model the behavior of the antennas in the vertical plane...}
%In the sequel, the model of the considered system in this paper is expressed in details. Meanwhile, our adopted notations are represented in Table \ref{tabs}.

%
%%%%%%%%%%%%%%%%%%%%%%%%%%%%%%%%%%%%%%%%%%%%%%%%%%%%%%%%%%%%%%%%
%%%%%%%%%%%%%%%%%%%%%%%%%%%%%%%%%%%%%%%%%%%%%%%%%%%%%%%%%%%%%%%% VERSUS P_T
%%%%%%%%%%%%%%%%%%%%%%%%%%%%%%%%%%%%%%%%%%%%%%%%%%%%%%%%%%%%%%%%
%%%%%%%%%%%%%%%%%%%%%%%%%%%%%%%%%%%%%%%%%%%%%%%%%%%%%%%%%%%%%%%%
%
%
%

%%%%%%%%%%%%%%%%%%%%%%%%%%%%%%%%%%%%%%%%%%%%%%%%%%%%%%%%%%%%%%%%
%%%%%%%%%%%%%%%%%%%%%%%%%%%%%%%%%%%%%%%%%%%%%%%%%%%%%%%%%%%%%%%% VERSUS P_T
%%%%%%%%%%%%%%%%%%%%%%%%%%%%%%%%%%%%%%%%%%%%%%%%%%%%%%%%%%%%%%%%
%%%%%%%%%%%%%%%%%%%%%%%%%%%%%%%%%%%%%%%%%%%%%%%%%%%%%%%%%%%%%%%%
%
%

%%%%%%%%%%%%%%%%%%%%%%%%%%%%%%%%%%%%%%%%%%%%%%%%%%%%%%%%%%%%%%%%
%%%%%%%%%%%%%%%%%%%%%%%%%%%%%%%%%%%%%%%%%%%%%%%%%%%%%%%%%%%%%%%% VERSUS W_Z
\begin{figure*}
	\centering
	\subfloat[] {\includegraphics[width=2.3 in]{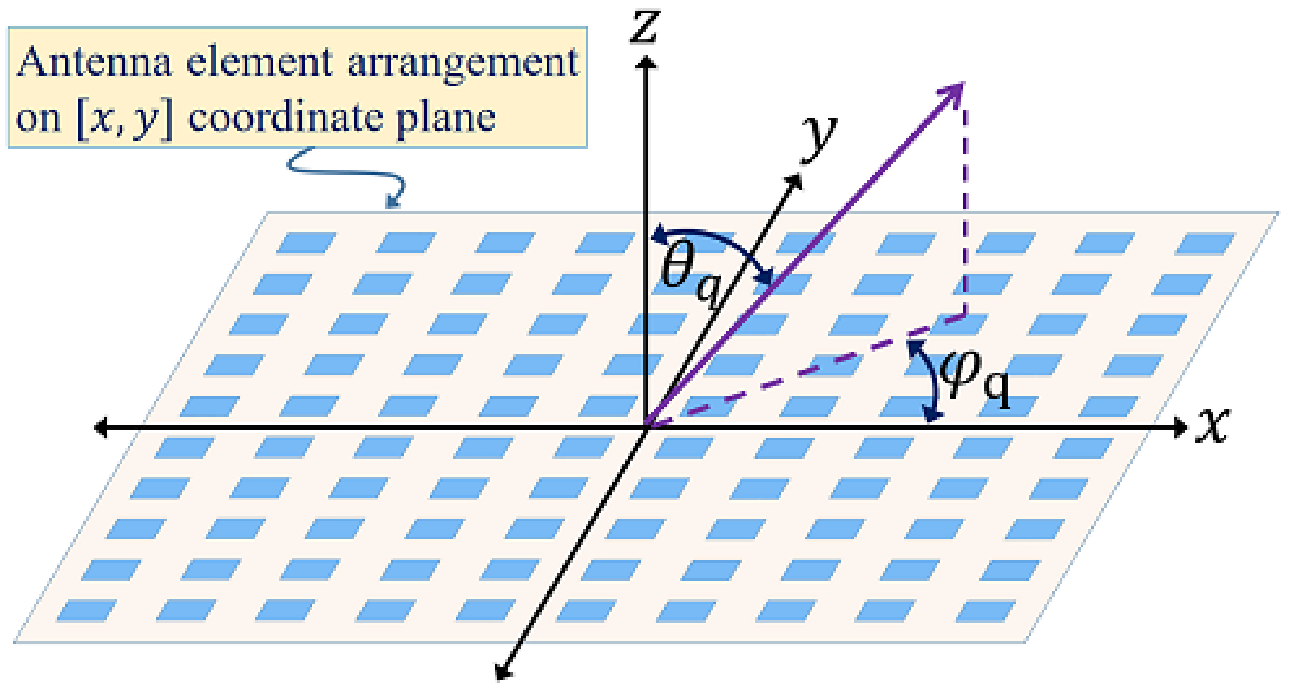}
		\label{x1}
	}
	\subfloat[] {\includegraphics[width=2.3 in]{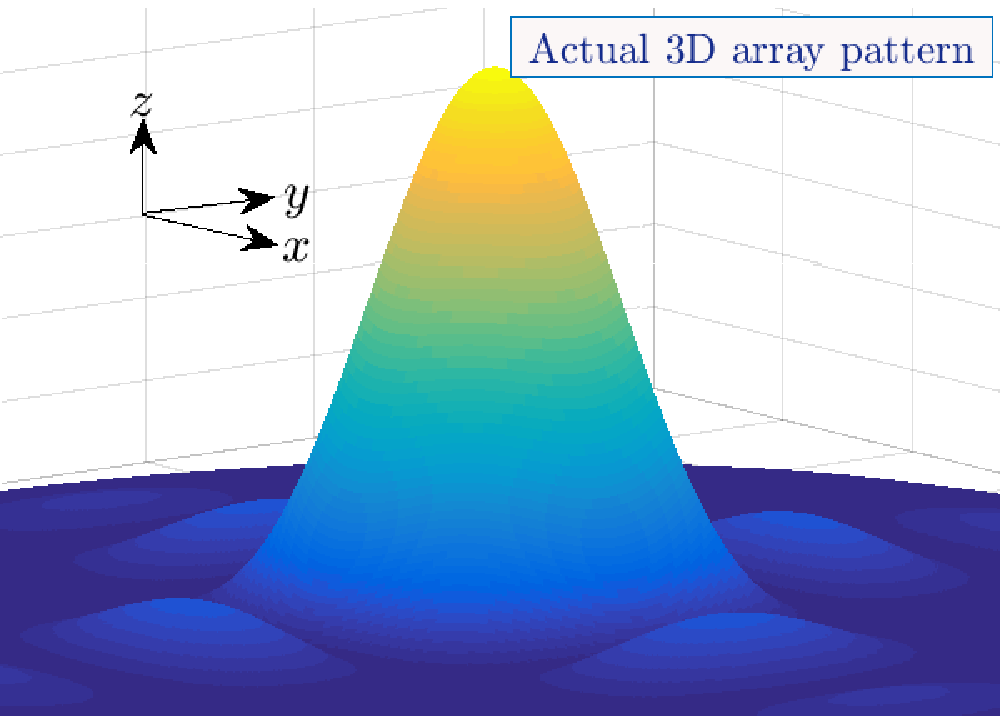}
		\label{x2}
	}
	\subfloat[] {\includegraphics[width=2.3 in]{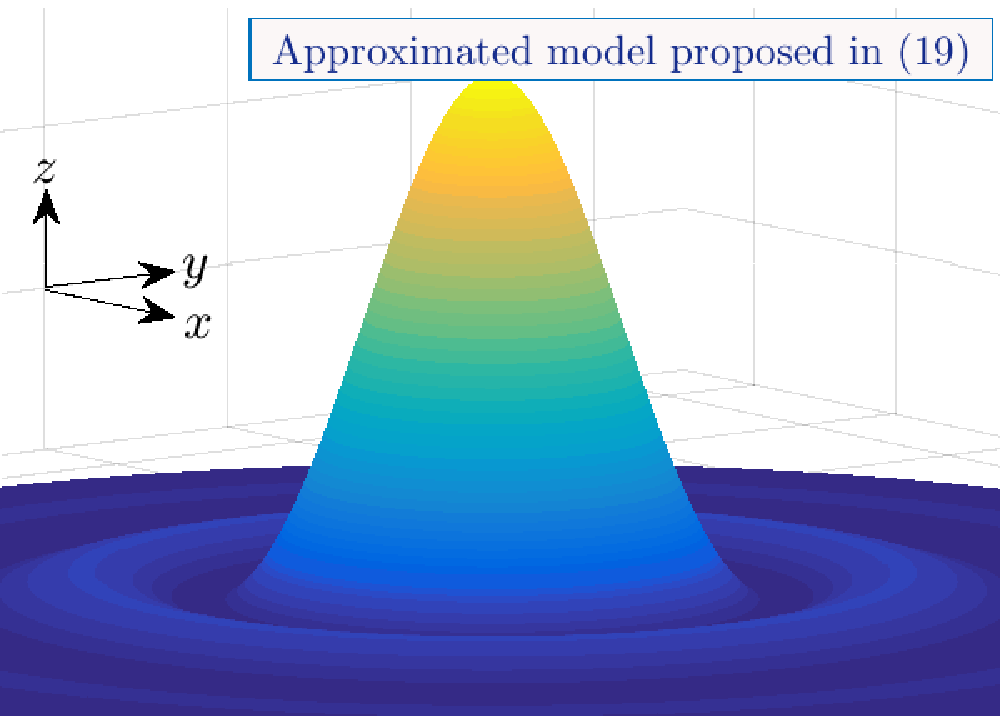}
		\label{x3}
	}\caption{3D illustration antenna pattern generated by a uniform $N_{q}\times N_{q}$ antenna array: (a) showing a $N_{q}\times N_{q}$ square array antenna arranged in $x-y$ plane; (b) 3D actual antenna pattern generated by a square array antenna arranged on the $x-y$ plane; (c) approximated antenna pattern obtained by \eqref{xs}.}
	\label{x4}
\end{figure*}
%%%%%%%%%%%%%%%%%%%%%%%%%%%%%%%%%%%%%%%%%%%%%%%%%%%%%%%%%%%%%%%%
%%%%%%%%%%%%%%%%%%%%%%%%%%%%%%%%%%%%%%%%%%%%%%%%%%%%%%%%%%%%%%%%

%-----------------------------------------
%-----------------------------------------
\subsection{3D Actual Antenna Pattern}
%-----------------------------------------
%-----------------------------------------
%This part captures the antenna array structures considered in this paper. 
Due to an approximate symmetry in the UAV vibrations in the $x$- and $y$-direction, as illustrated in Fig. \ref{x1}, we consider a uniform square array antenna, comprising $N_q\times N_q$ antenna elements with the same spacing between elements in $x$- and $y$-direction, i.e., $d_x=d_y=d_a$ where $d_x$ and $d_y$ are spacing between antenna elements in x- and $y$-direction, respectively.
%------------------
The array radiation gain is mainly formulated in the direction of $\theta_{q}$ and $\phi_{q}$. In our model, $\theta_{q}$ and $\phi_{q}$ can be defined as  functions of random variables (RVs) $\theta_{qx}$ and $\theta_{qy}$ as follows:
\begin{align}
\label{f_2}
\theta_{q}  &= \tan^{-1}\left(\sqrt{\sin^2(\theta_{qx})+\sin^2(\theta_{qy})}\right), \nonumber \\
\phi_q &=\tan^{-1}\left(\frac{\sin(\theta_{qy})}{\sin(\theta_{qx})}\right).
\end{align} 
By taking into account the effect of all elements, the array radiation gain in direction of angles $\theta_{qx}$ and $\theta_{qy}$ will be:
\begin{align}
\label{p_1}
G_{q}(\theta_{qx},\theta_{qy})  = G_a(\theta_{qx},\theta_{qy})\,G_e(\theta_{qx},\theta_{qy}),
\end{align}
where $G_a$ is an array factor and $G_e$ is single element radiation pattern. From the 3GPP single element radiation pattern, $G_{e,\textrm{3dB}}=10\times\log_{10}(G_e)$ of each single antenna element is obtained as \cite{niu2015survey}
\begin{align}
	\label{vb1}
	&G_{e\textrm{3dB}} = G_{\textrm{max}} - \min\left\{-(G_{e\textrm{3dB,1}}+G_{e\textrm{3dB,2}}),F_m 
	\right\},  \\
	%-----------------------------
	&G_{e\textrm{3dB,1}} =  -\min \left\{ - 12\left(\frac{\theta_e-90}{\theta_{e\textrm{3dB}}}\right)^2,
	G_{\textrm{SL}}\right\},      
	\nonumber \\
	%-----------------------------
	&G_{e\textrm{3dB,2}} = -\min \left\{ - 12\left(\frac{\theta_{qx}}{\phi_{e\textrm{3dB}}}\right)^2,
	F_m\right\},
	\nonumber \\
	%-----------------------------
	&\theta_e             = \tan^{-1}\left( \frac{\sqrt{1+\sin^2(\theta_{qx})}}
	{\sin(\theta_{qy})} \right),
	\nonumber 
\end{align}
where $\theta_{e\textrm{3dB}}=65^{\circ}$ and $\phi_{e\textrm{3dB}}=65^{\circ}$ are the vertical and horizontal 3D beamwidths, respectively, $G_{\textrm{max}}=8$ dBi is the maximum directional gain of the antenna element, $F_m=30$ dB is the front-back ratio, and $G_{\textrm{SL}}=30$ dB is the side-lobe level limit. 

If the amplitude excitation of the entire array is uniform, then the array factor $G_a(\theta_{qx},\theta_{qy})$ for a square array of $N_{q}^2$ elements can be obtained as \cite{balanis2016antenna}
\begin{align}
\label{f_1}
G_a(\theta_{qx},\theta_{qy}) &= G_0(N_{q})
\left( \frac{\sin\left(\frac{N_{q} (k d_x \sin(\theta_{q})\cos(\phi_{q})+\beta_x)}{2}\right)} 
{N_{q}\sin\left(\frac{k d_x \sin(\theta_{q})\cos(\phi_{q})+\beta_x}{2}\right)}
\right. \nonumber\\
%------------
&~~~~~\times \left. \frac{\sin\left(\frac{N_{q} (k d_y \sin(\theta_{q})\sin(\phi_{q})+\beta_y)}{2}\right)} 
{N_{q}\sin\left(\frac{k d_y \sin(\theta_{q})\sin(\phi_{q})+\beta_y}{2}\right)}\right)^2,
\end{align}
where $d_x=\frac{\lambda}{2}$ and $\beta_x$ are the spacing and progressive phase shift between the elements along the  $x$ axis, respectively, and $d_y=\frac{\lambda}{2}$ and $\beta_y$ are the spacing and progressive phase shift between the elements along the $y$ axis, respectively, $k=\frac{2\pi}{\lambda}$ denotes the wave number, $\lambda=\frac{c}{f_c}$ denotes wavelength, $f_c$ denotes the carrier frequency and $c$ iss the speed of light. 
%--------
One of our key goals is to answer this question that for a UAV with a given instability, i.e., a given standard deviation of orientation fluctuations $\sigma_{to}$, what is the optimum values of $N_q$ that achieves minimum outage probability. 
%
%One of our key goals is to answer this question that for a UAV with a given stability, what is the optimum values of $N_\textrm{q}$ that achieves the best performance. 
%
Hence, for a fair comparison between antennas with different $N_{q}$, we consider the total radiated power of antennas with different $N_{q}$ are same. From this, we have
\begin{align}
\label{cv}
G_0(N_{q})=\frac{G'_0}{\int_0^{\pi}\int_0^{2\pi} G(\theta_{q},\phi_{q}) \sin(\theta_{q}) d\theta_{q} d\phi_{q}}.
\end{align}
%For simplicity and without loss of generality, we consider $G'_0=1$.
%--------
More details on the element and array radiation pattern is provided in \cite{niu2015survey,balanis2016antenna}.
%--------
%As depicted in Fig. \ref{st1}, $z$ axis refers to the direction that extends from Tx (Tx node placed in [0,0,0] Cartesian coordinate) toward Rx node (Rx node placed on [0,0,$z$]). 
%--------
In addition, without loss of generality, it is assumed that $\beta_x=\beta_y=0$ and the hovering UAV sets its antenna main-lobe direction on $z$ axis. 
%
%Now, due to the orientation fluctuations of hovering UAVs, direction of antenna mounted on UAV fluctuates around the $z$ axis as $\theta_q=[\theta_\textrm{qx},\theta_\textrm{qy}]$ depicted in Fig. \ref{}.
%--------
Now, the instantaneous directivity gain of a A2A link will be given by \cite{yu2017coverage}
\begin{align}
\label{pk1}
\mathbb{G}_{\textrm{uu}}(\theta_{{tx}},\theta_{{ty}},\theta_{{rx}},\theta_{{ry}}) 
= G_t(\theta_{{tx}},\theta_{{ty}}) 
\times G_r(\theta_{{rx}},\theta_{{ry}}).
\end{align}
From \eqref{pk1}, we can see that the instantaneous directivity is a function of four independent RVs $\theta_{{tx}},\theta_{{ty}},\theta_{{rx}},\theta_{{ry}}$.

%++++++++++++++++++++++++Planar arrays provide additional variables which can be used to control and shape the pattern of the array. Planar arrays are more versatile and can provide more symmetrical patterns with lower side lobes. In addition, they can be used to scan the main beam of the antenna toward any point in space.

%++++=the 3D antenna element gain pattern

%+++For a fair comparison, we consider the power excitation of the entire array is uniform as $P_{\textrm{te}}=\frac{P_t}{N^2}$

%-----------------------------------------------------------
%-----------------------------------------------------------
\subsection{Received Signal Model}
%-----------------------------------------------------------
%----------------------------------------------------------- 
Given the 3D antenna pattern described in the previous subsection, the end-to-end signal-to-interference-plus-noise ratio (SINR) of a A2A link can be obtained as
\begin{align}
\label{s1}
\gamma_{\textrm{uu}}(\alpha,\theta_{{tx}},\theta_{{ty}},\theta_{{rx}},\theta_{{ry}})  \!=\! \frac{P_t|\alpha|^2 h_L(Z) \mathbb{G}(\theta_{{tx}},\theta_{{ty}},\theta_{{rx}},\theta_{{ry}})}
{ \Sigma I + \sigma^2}\!, 
\end{align}
where $\sigma^2$ is the thermal noise power, $\alpha$ is the small scale fading coefficient, $P_t$ is the transmit power, and $h_L(Z)$ is the path loss coefficient.
%%%%%%%%%%%%%%%%
Moreover, in \eqref{s1}, $\Sigma I=\Sigma I_{\textrm{d}}+\Sigma I_{{r}}$, where $\Sigma I_{\textrm{d}}$ and $\Sigma I_{{r}}$ are the inter-carrier interference due to Doppler spread, and radio interference due to the other Txs, respectively.
%%%%%%%%%%%%%%%%
Note that, by using high directional radio patterns at the Rx, $\Sigma I_{{r}}$ can be effectively eliminated \cite{lin2018sky,3gpp1}.
%the probability that transmitted signals by other transmitters with the same frequency of considered transmitter  outside of the main lobe of the UAV antenna
%%%%%%%%%%%%%%%%
Furthermore, $\Sigma I_{\textrm{d}}$ is caused by Doppler spread and it is proportional to $\Sigma I_{\textrm{d}} \propto \left[1-\textrm{sinc}^2(f_{\textrm{d}}T_s)\right]$, where $T_s$ is the symbol duration, $f_{\textrm{d}} = \frac{f_c \nu}{c}$ is the Doppler frequency shift, $c=3\times 10^8$ (in m/s) is the speed of light, $\nu$ (in m/s) is the relative moving velocity, and $f_c$ (in GHz) is the carrier frequency \cite{sesia2011lte}.
Moreover, in \cite{zhou2019beam}, it was shown that for a moving UAV with $\nu\leq 10$\,m/s,  the impact of the Doppler spread is negligible. In our setup, we assume that UAVs are hovering at a fixed position, i.e., multi-rotor UAVs or tethered balloons, and there is no relative velocity between communication nodes; therefore, there will be no Doppler spreading effect. As a result, expression in \eqref{s1} can be simplified to
\begin{align}
\label{ss1}
\gamma_{\textrm{uu}}(\alpha,\theta_{{tx}},\theta_{{ty}},\theta_{{rx}},\theta_{{ry}})  \!=\! 
\frac{P_t|\alpha|^2 h_L(Z) \mathbb{G}(\theta_{{tx}},\theta_{{ty}},\theta_{{rx}},\theta_{{ry}})}
{  \sigma^2}.
\end{align}
%%%%%%%%%%%%%%%%

Since there is still no standardized results for UAV-based communications at mmW bands, we consider the results of the recent 3GPP report in \cite{3gppf} in order to set the path loss parameters. These parameters are valid for a BS height up to  150 m and are expressed as follows:
\begin{align}
\label{loss}
h_{\textrm{L,dB}}(Z) &= -20 \log_{10}\left(\frac{40 \pi Z f_c}{3}\right)  \\
&+ \min\left\{0.03 h_b ^{1.73},10\right\}\times \log_{10}(Z) \nonumber \\
%%%%%%%%
& + \min\left\{0.044 h_b ^{1.73},14.77\right\}
-0.002\, Z \log_{10}(h_b),  \nonumber
\end{align}
where  $h_b$ (in meter) is the average of building height of city.
%%------------------------------------------------------
%%------------------------------------------------------
Moreover, from the measurement results provided in \cite{goddemeier2015investigation}, for a low altitude communication link between UAVs, Rician and Nakagami distributions were shown to be highly promising models  that can be mathematically fitted into the experimentally measured data.
Since the Nakagami distribution is a universal model that can capture various channel conditions, we apply it to model small-scale fading.
Let $\alpha$ be a Nakagami random variable (RV). Hence, $\zeta= \alpha^2$ will be a normalized Gamma RV given by:
\begin{align}
\label{Gamma}
f_\zeta (\zeta) = \frac{m^m  \zeta^{m-1}}{\Gamma(m)}     \exp(-m \zeta), ~~~\zeta>0,
\end{align}
where $m$ is the Nakagami fading parameter and $\Gamma(\cdot)$ is the Gamma function \cite{goddemeier2015investigation}.

In practice, a highly directional beam is used to compensate the high free-space path loss at the mmW band.  Hence, in addition to the channel fading, fluctuations in the orientation of the UAVs (due to the effect of wind, mechanical and control system flaws, antenna and BS payload, etc.) can  lead to beam misalignment and adversely affect the link performance and channel capacity. To capture these effects, we define the outage capacity, i.e., the probability with which the instantaneous capacity falls bellow a certain threshold $\mathcal{C}_{\textrm{th}}$, as the figure of merit to determine the reliability of the considered UAV-based communication system. The outage capacity can be defined as follows:
\begin{align}
\label{xd}
\mathbb{P}_{\textrm{out}} &= {\textrm{Pr}}\{\log_2(1+\gamma)<\mathcal{C}_{\textrm{th}}\}
= F_{\gamma}(\gamma_{\textrm{th}}),
\end{align}
where $F_x(\cdot)$ is the cumulative distribution function (CDF) of RV $x$, $\gamma$ is the end-to-end signal-to-noise ratio (SNR), and $\gamma_{\textrm{th}}=2^{\mathcal{C}_{\textrm{th}}}-1$ is the SNR threshold.

From \eqref{ss1}, it can be observed that the end-to-end SNR is composed of the deterministic loss parameter $h_L$, and several RVs, i.e., the small-scale fading coefficient $\alpha$, the AoD deviations due to $\theta_{{tx}}$ and  $\theta_{{ty}}$, and the AoA deviations $\theta_{{rx}}$ and $\theta_{{ry}}$. 
%-----
To assess the benefits of deploying UAV-based mmW communications under the aforementioned RVs, one important challenge is to accurately model the channel, which can then be used for easily evaluating the performance of hovering UAV-based mmW links without performing time-consuming simulations.
%-----
Accordingly, in the next section, we derive the closed-form expressions of the SNR distribution at the Rx by taking into account the unique characteristics of mmW links along with the effects of UAV random vibrations and orientation fluctuations.

%%%%%%%%%%%%%%%%%%%%%%%%%%%%%%%%%%%%%%%%%%%%%%%%%%%%%%%%%%%%%%%%%%
%%%%%%%%%%%%%%%%%%%%%%%%%%%%%%%%%%%%%%%%%%%%%%%%%%%%%%%%%%%%%%%%%%
\section{Analytical Channel Models}
%%%%%%%%%%%%%%%%%%%%%%%%%%%%%%%%%%%%%%%%%%%%%%%%%%%%%%%%%%%%%%%%%%
%%%%%%%%%%%%%%%%%%%%%%%%%%%%%%%%%%%%%%%%%%%%%%%%%%%%%%%%%%%%%%%%%%
Next, we first develop a channel model for the A2A link. Then, for simpler A2G and G2A cases, we obtain  more tractable channel models.

%%%%%%%%%%%%%%%%%%%%%%%%%%% begin THEOREM 1 %%%%%%%%%%%%%%%%%%%%%%%%%%%%%%%%%%%%%%%%%%%%%%%%%%%%%%%%%%%%%%%%%%%%%%%%%%%%%%%%%%%%%%%%%%%%%%%%%%%%%%%%%%%%%%%%%%%%%%%%%%
%%%%%%%%%%%%%%%%%%%%%%%%%%% begin THEOREM 1 %%%%%%%%%%%%%%%%%%%%%%%%%%%%%%%%%%%%%%%%%%%%%%%%%%%%%%%%%%%%%%%%%%%%%%%%%%%%%%%%%%%%%%%%%%%%%%%%%%%%%%%%%%%%%%%%%%%%%%%%%%
%%%%%%%%%%%%%%%%%%%%%%%%%%% begin THEOREM 1 %%%%%%%%%%%%%%%%%%%%%%%%%%%%%%%%%%%%%%%%%%%%%%%%%%%%%%%%%%%%%%%%%%%%%%%%%%%%%%%%%%%%%%%%%%%%%%%%%%%%%%%%%%%%%%%%%%%%%%%%%%

\subsection{UAV-to-UAV Link}
%\begin{thm}
%{\textbf{\textsl{Theorem 1}}} {\it (PDF and CDF of of instantaneous SNR of U2U link):}
{\bf Theorem 1.}
%{\it(PDF and CDF of instantaneous SNR for U2U link):}
{\it The probability density function (PDF) of end-to-end SNR of A2A link can be well modeled as
\begin{align}
    \label{uu}
	&f_{\gamma_{\textrm{uu}}}(\gamma_{\textrm{uu}}) =\! \sum_{d_{r}=0}^{jD-1} \sum_{d_{t}=0}^{jD-1}\!
	\mathbb{R}_{d_{t},d_{r}}
	%---------------------------
	\gamma_{\textrm{uu}}^{m-1}     
	\exp\!\left(\!- \frac{m \sigma^2}{P_t h_L(Z) \mathbb{R}'_{d_{t},d_{r}} }\,\gamma_{\textrm{uu}}\! \right)\!,
\end{align}
where for $d_{t}\,\&\, d_{r} \in \{0,1,...,jD-1\}$ and $j\in\{1,2\}$, we have
\begin{align}
\mathbb{R}_{d_{t},d_{r}} = \frac{\mathbb{J}_{d_{t},d_{r}}(\theta'_{t,xy},\theta'_{r,xy},\sigma^2_{to},\sigma^2_{ro})}{\Gamma(m) }
\left(\frac{m\,\sigma^2}{P_t h_L(Z) \mathbb{R}'_{d_{t},d_{r}} }\right)^m, \nonumber
\end{align}
%
%
%%%%%%%%%%%%%%%%%%%%%%%%%%%%%%%%%%%
%
%
\begin{equation}
\left \{
\begin{array}{ll}
\mathbb{R}'_{0,0} &\!\!\!\!\!\!= 4 k^4 d_a^4 G_0''(N_t) G_0''(N_r), \\
%---------------------------
\mathbb{R}'_{d_{t},0} &\!\!\!\!\!\!= 2 k^2 d_a^2 G_0''(N_t)
G_0''(N_r) \frac{D^2\left(1-\cos\left( \frac{d_{t} k d_a}{D}\right)\right)} {d_{t}^2},\\
%---------------------------
\mathbb{R}'_{0,d_{r}} &\!\!\!\!\!\!= 2 k^2 d_a^2 G_0''(N_t) 
G_0''(N_{r}) \frac{D^2\left(1-\cos\left( \frac{d_{r} k d_a}{D}\right)\right)} {d_{r}^2},\\
%---------------------------
\mathbb{R}'_{d_{t},d_{r}} &\!\!\!\!\!\!= 4G_0''(N_t) G_0''(N_{r}) 
\frac{D^4\left(\sin^2\left( \frac{d_{t} k d_a}{2 D}\right)   \sin^2\left( \frac{d_{r} k d_a}{2 D}\right)   \right)} 
{d_{t}^2\,d_{r}^2},
\end{array}
\right. \nonumber
\end{equation}
%
%
%%%%%%%%%%%%%%%%%%%%%%%%%%%%%%%%%%%
%
%
\begin{align}
\mathbb{J}_{d_{t},d_{r}}(\theta'_{t,xy},\theta'_{r,xy},\sigma^2_{to},\sigma^2_{ro}) 
= J_{d_{t}}(\theta'_{t,xy},\sigma^2_{to}) J_{d_{r}}(\theta'_{r,xy},\sigma^2_{ro}), \nonumber
\end{align}
%
%
%%%%%%%%%%%%%%%%%%%%%%%%%%%%%%%%%
%
%
\begin{align}
& J_{d_{q}}(\theta'_{q,xy},\sigma^2_{qo}) =
M\!\left(\frac{\theta'_{q,xy}}{\sigma_{qo}},\frac{d_{q}}{D N_{q} \sigma_{qo}}\right)
\!-\! M\left(\!\frac{\theta'_{q,xy}}{\sigma_{qo}},\frac{d_{q}+1}{D N_{q} \sigma_{qo}}\!\right)\!. \nonumber
\end{align}
%
%
%%%%%%%%%%%%%%%%%%%%%%%%%%%%%%%%%%%
%
%
In addition, $G_0''(N_{q})=0.2025\times 10^{\frac{G_{\textrm{max}}}{10}}G_0(N_{q})$ and $M(a,b)$ is the Marcum {\it Q}-function.
}
%\end{thm}

%For instance, in the proposed analytical results, antenna directivity gain and beamwidth are tuned by parameter $N$ that by increasing $N$ antenna directivity gain increases with the decreasing beamwidth. Also, the severity of UAV instabilities will be tuned by $\sigma_{\textrm{ty}}$, $\sigma_{\textrm{ry}}$, $\theta'_{\textrm{ty}}$ and $\theta'_{\textrm{ry}}$, and small-scale fading strength will be tuned by $m$.
%%%%%%%%%%%%%%%%%%

%As we observe, the given analytical expressions for channel distribution and outage probability compose the effects of large- and small-scale fading, UAVs fluctuations, and antenna pattern specification in the closed-form expressions. Moreover, the given expressions are based on the standard built-in functions which is available in popular mathematical software packages, e.g., MATLAB, and Mathematica.
%%%%%
%Therefore, the given tractable analytical channel distribution will be suitable to analytically analyze important performance metrics such as channel capacity and bit error rate for academia and industry studies without requiring time-consuming simulations.

%%%%%%%%%%%%%%%%%%%%%%%%%%%%%%%%%%%%%%%%%%%%%%%%%%%%%%%%%%%%%%%%
%%%%%%%%%%%%%%%%%%%%%%%%%%%%%%%%%%%%%%%%%%%%%%%%%%%%%%%%%%%%%%%% VERSUS W_Z
\begin{figure*}
	\centering
	\subfloat[] {\includegraphics[width=2.1 in]{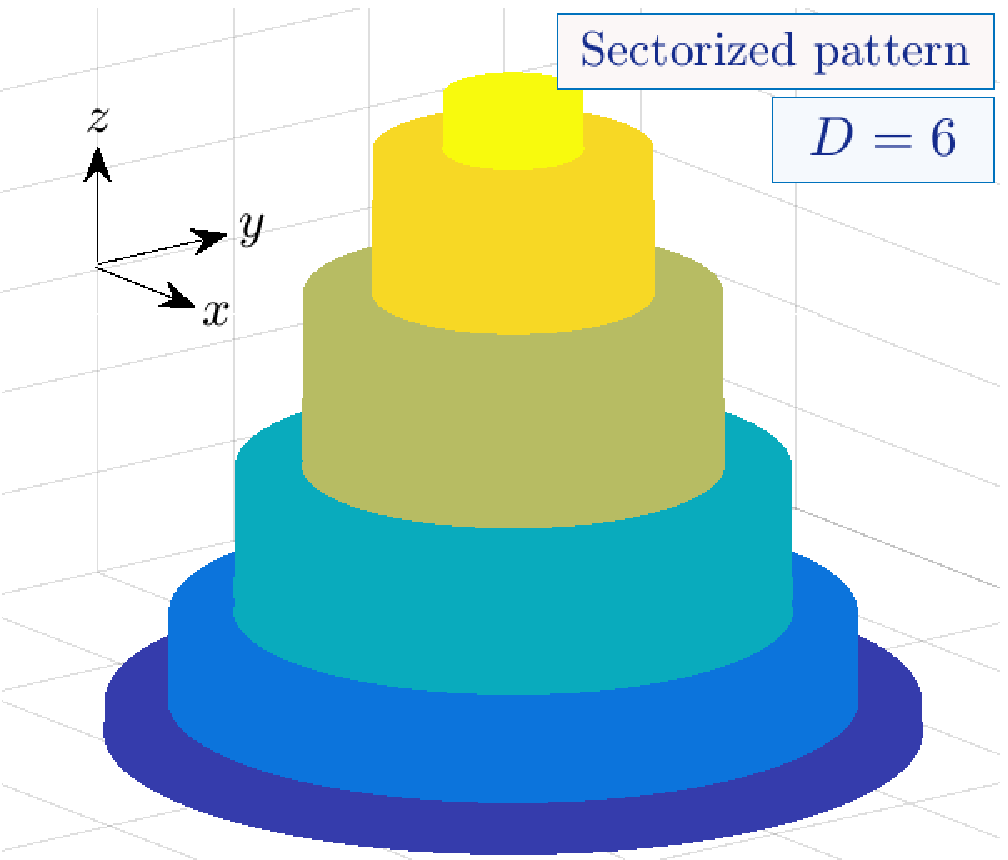}
		\label{xx2}
	}
	%\hfill
	\subfloat[] {\includegraphics[width=2.1 in]{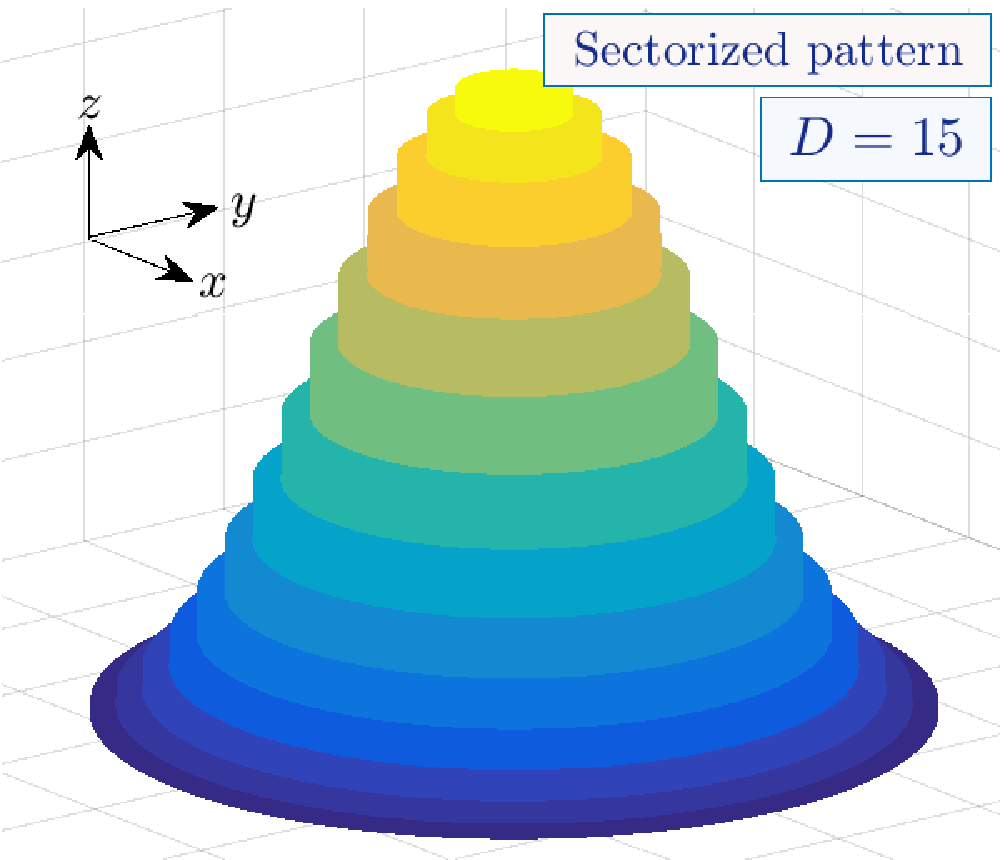}
		\label{xx3}
	}
	\subfloat[] {\includegraphics[width=2.7 in]{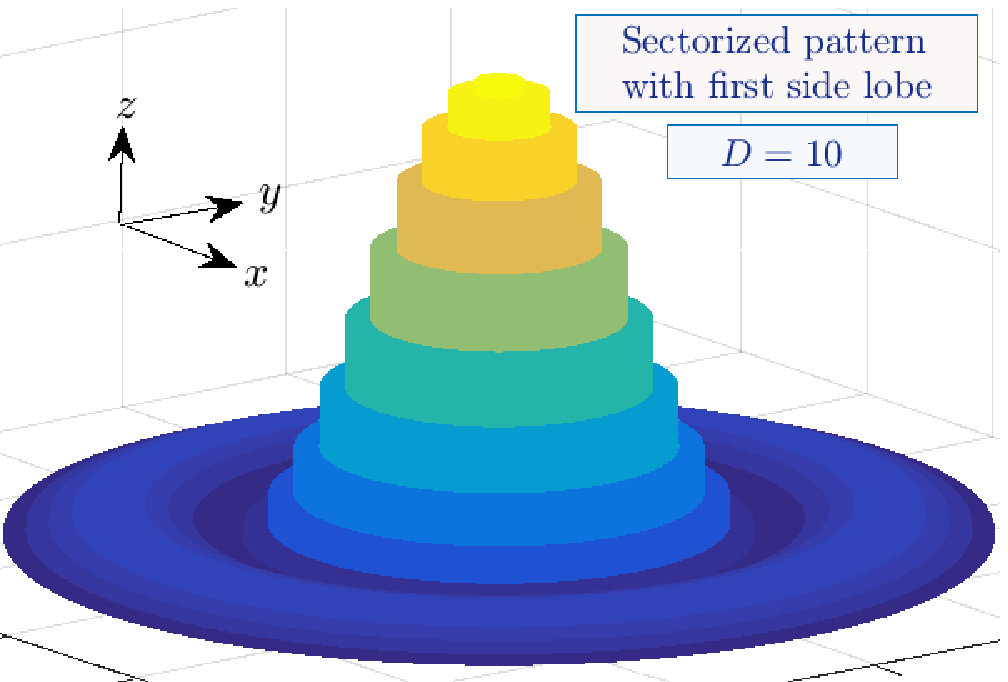}
	    \label{xx4}
    }
	\caption{Sectorized antenna pattern obtained from \eqref{kgg} for (a) $D=6$ and $j=1$, (b) $D=15$ and $j=1$, and (c) $D=10$ and $j=2$. }
	\label{x5}
\end{figure*}
%%%%%%%%%%%%%%%%%%%%%%%%%%%%%%%%%%%%%%%%%%%%%%%%%%%%%%%%%%%%%%%%
%%%%%%%%%%%%%%%%%%%%%%%%%%%%%%%%%%%%%%%%%%%%%%%%%%%%%%%%%%%%%%%%

%%%%%%%%%%%%%%%%%%%%%%%%%%% begin PROOF %%%%%%%%%%%%%%%%%%%%%%%%%%%%%%%%%%%%%%%%%%%%%%%%%%%%%%%%%%%%%%%%%%%%%%%%%%%%%%%%%%%%%%%%%%%%%%%%%%%%%%%%%%%%%%%%%%%%%%%%%%
%%%%%%%%%%%%%%%%%%%%%%%%%%% begin PROOF %%%%%%%%%%%%%%%%%%%%%%%%%%%%%%%%%%%%%%%%%%%%%%%%%%%%%%%%%%%%%%%%%%%%%%%%%%%%%%%%%%%%%%%%%%%%%%%%%%%%%%%%%%%%%%%%%%%%%%%%%%
%%%%%%%%%%%%%%%%%%%%%%%%%%% begin PROOF %%%%%%%%%%%%%%%%%%%%%%%%%%%%%%%%%%%%%%%%%%%%%%%%%%%%%%%%%%%%%%%%%%%%%%%%%%%%%%%%%%%%%%%%%%%%%%%%%%%%%%%%%%%%%%%%%%%%%%%%%%
%%%%%%%%%%%%%%%%%%%%%%%%%%% begin PROOF %%%%%%%%%%%%%%%%%%%%%%%%%%%%%%%%%%%%%%%%%%%%%%%%%%%%%%%%%%%%%%%%%%%%%%%%%%%%%%%%%%%%%%%%%%%%%%%%%%%%%%%%%%%%%%%%%%%%%%%%%%
%%%%%%%%%%%%%%%%%%%%%%%%%%% begin PROOF %%%%%%%%%%%%%%%%%%%%%%%%%%%%%%%%%%%%%%%%%%%%%%%%%%%%%%%%%%%%%%%%%%%%%%%%%%%%%%%%%%%%%%%%%%%%%%%%%%%%%%%%%%%%%%%%%%%%%%%%%%
\begin{IEEEproof}
Please refer to Appendix \ref{AppA}.
\end{IEEEproof}
%%%%%%%%%%%%%%%%%%%%%%%%%%% END PROOF %%%%%%%%%%%%%%%%%%%%%%%%%%%%%%%%%%%%%%%%%%%%%%%%%%%%%%%%%%%%%%%%%%%%%%%%%%%%%%%%%%%%%%%%%%%%%%%%%%%%%%%%%%%%%%%%%%%%%%%%%%%%
%%%%%%%%%%%%%%%%%%%%%%%%%%% END PROOF %%%%%%%%%%%%%%%%%%%%%%%%%%%%%%%%%%%%%%%%%%%%%%%%%%%%%%%%%%%%%%%%%%%%%%%%%%%%%%%%%%%%%%%%%%%%%%%%%%%%%%%%%%%%%%%%%%%%%%%%%%%%
%%%%%%%%%%%%%%%%%%%%%%%%%%% END PROOF %%%%%%%%%%%%%%%%%%%%%%%%%%%%%%%%%%%%%%%%%%%%%%%%%%%%%%%%%%%%%%%%%%%%%%%%%%%%%%%%%%%%%%%%%%%%%%%%%%%%%%%%%%%%%%%%%%%%%%%%%%%%
%%%%%%%%%%%%%%%%%%%%%%%%%%% END PROOF %%%%%%%%%%%%%%%%%%%%%%%%%%%%%%%%%%%%%%%%%%%%%%%%%%%%%%%%%%%%%%%%%%%%%%%%%%%%%%%%%%%%%%%%%%%%%%%%%%%%%%%%%%%%%%%%%%%%%%%%%%%%
%%%%%%%%%%%%%%%%%%%%%%%%%%% END PROOF %%%%%%%%%%%%%%%%%%%%%%%%%%%%%%%%%%%%%%%%%%%%%%%%%%%%%%%%%%%%%%%%%%%%%%%%%%%%%%%%%%%%%%%%%%%%%%%%%%%%%%%%%%%%%%%%%%%%%%%%%%%%
%%%%%%%%%%%%%%%%%%%%%%%%%%% END PROOF %%%%%%%%%%%%%%%%%%%%%%%%%%%%%%%%%%%%%%%%%%%%%%%%%%%%%%%%%%%%%%%%%%%%%%%%%%%%%%%%%%%%%%%%%%%%%%%%%%%%%%%%%%%%%%%%%%%%%%%%%%%%
%%%%%%%%%%%%%%%%%%%%%%%%%%% END PROOF %%%%%%%%%%%%%%%%%%%%%%%%%%%%%%%%%%%%%%%%%%%%%%%%%%%%%%%%%%%%%%%%%%%%%%%%%%%%%%%%%%%%%%%%%%%%%%%%%%%%%%%%%%%%%%%%%%%%%%%%%%%%
%\textcolor{white}{p}

As one can observe from \eqref{uu}, the effects of large- and small-scale fading characterized by $\alpha$, $m$ and $h_L(Z)$, UAVs stability characterized by $\theta'_{tx}$, $\theta'_{ty}$, $\theta'_{rx}$, $\theta'_{ry}$, $\sigma^2_{to}$ and $\sigma^2_{ro}$, and antenna pattern specifications characterized by $N_t$ and $N_r$, are incorporated into the closed-form channel PDF. For instance, in the proposed analytical results, antenna directivity gain and beamwidth are tuned by $N_{q}$ that by increasing $N_{q}$ antenna directivity gain increases with the decreasing beamwidth. In addition, the strength of UAV instability is modeled by $\sigma_{{to}}$, $\sigma_{{ro}}$, $\theta'_{{tx}}$, $\theta'_{{ty}}$, $\theta'_{{rx}}$ and $\theta'_{{ry}}$, and the small-scale fading strength is characterized by $m$.

We note that the accuracy of the derived analytical expressions in Theorem 1 depends on the variable $D$ that is used for the approximation of antenna pattern. The optimal value of $D$ is its minimum value that satisfies a predefined accuracy. As seen later, $D=30$ is a good choice and achieves the analytical results close to the simulation results.

Moreover, note that parameter $j\in\{1,2\}$ is another parameter that controls the accuracy of the derived channel PDF. 
%The value $j=1$ is used when the main-lobe of pattern is considered and for more precision, $j=2$ is used when main-lobe along with the first side-lobe is considered. 
The accuracy of the derived channel model for both values of $j$ are investigated in the Section \textcolor{black}{III} using simulation results. 

Now we derive the analytical expression for CDF of end-to-end SNR of A2A link which is useful for outage probability analysis.

%%%%%%%%%%%%%%%%%%%%%%%%%%% begin LEMMA 1 %%%%%%%%%%%%%%%%%%%%%%%%%%%%%%%%%%%%%%%%%%%%%%%%%%%%%%%%%%%%%%%%%%%%%%%%%%%%%%%%%%%%%%%%%%%%%%%%%%%%%%%%%%%%%%%%%%%%%%%%%%
%%%%%%%%%%%%%%%%%%%%%%%%%%% begin LEMMA 1 %%%%%%%%%%%%%%%%%%%%%%%%%%%%%%%%%%%%%%%%%%%%%%%%%%%%%%%%%%%%%%%%%%%%%%%%%%%%%%%%%%%%%%%%%%%%%%%%%%%%%%%%%%%%%%%%%%%%%%%%%%
%%%%%%%%%%%%%%%%%%%%%%%%%%% begin LEMMA 1 %%%%%%%%%%%%%%%%%%%%%%%%%%%%%%%%%%%%%%%%%%%%%%%%%%%%%%%%%%%%%%%%%%%%%%%%%%%%%%%%%%%%%%%%%%%%%%%%%%%%%%%%%%%%%%%%%%%%%%%%%%
{\bf Lemma 1.} {\it The CDF of end-to-end SNR of A2A link is obtained as
\begin{align}
\label{cuu}
F_{\gamma_{\textrm{uu}}}(\gamma_{\textrm{uu}}) &=\sum_{d_{r}=0}^{jD-1} \sum_{d_{t}=0}^{jD-1}\!
\mathbb{R}_{d_{t},d_{r}}
\left(\frac{P_t h_L(Z) \mathbb{R}'_{d_{t},d_{r}} }{m \sigma^2}\right)^m\nonumber \\
%---------------------------  
&~~~~~~~~~~~~~~\times\mathcal{V}\left(m, \frac{m \sigma^2}{P_t h_L(Z) \mathbb{R}'_{d_{t},d_{r}} }\,\gamma_{\textrm{uu}} \right),
\end{align}
where $\mathcal{V}(.,.)$ is the incomplete Gamma function.
}

\begin{IEEEproof}
Using \eqref{uu} and \cite[(8.350.1)]{jeffrey2007table}, the CDF of RV $\gamma_{\textrm{uu}}$ is derived in \eqref{cuu}.
\end{IEEEproof}
%%%%%%%%%%%%%%%%%%%%%%%%%%% END LEMMA 1 %%%%%%%%%%%%%%%%%%%%%%%%%%%%%%%%%%%%%%%%%%%%%%%%%%%%%%%%%%%%%%%%%%%%%%%%%%%%%%%%%%%%%%%%%%%%%%%%%%%%%%%%%%%%%%%%%%%%%%%%%%
%%%%%%%%%%%%%%%%%%%%%%%%%%% END LEMMA 1 %%%%%%%%%%%%%%%%%%%%%%%%%%%%%%%%%%%%%%%%%%%%%%%%%%%%%%%%%%%%%%%%%%%%%%%%%%%%%%%%%%%%%%%%%%%%%%%%%%%%%%%%%%%%%%%%%%%%%%%%%%
%%%%%%%%%%%%%%%%%%%%%%%%%%% END LEMMA 1 %%%%%%%%%%%%%%%%%%%%%%%%%%%%%%%%%%%%%%%%%%%%%%%%%%%%%%%%%%%%%%%%%%%%%%%%%%%%%%%%%%%%%%%%%%%%%%%%%%%%%%%%%%%%%%%%%%%%%%%%%%

%%--------------------------------------------------
%%--------------------------------------------------
\subsection{Ground-to-UAV and UAV-to-Ground Links}
%%--------------------------------------------------
%%--------------------------------------------------
Next we derive the channel distribution of G2A and A2G links. 

%%%%%%%%%%%%%%%%%%%%%%%%%%% begin LEMMA 2 %%%%%%%%%%%%%%%%%%%%%%%%%%%%%%%%%%%%%%%%%%%%%%%%%%%%%%%%%%%%%%%%%%%%%%%%%%%%%%%%%%%%%%%%%%%%%%%%%%%%%%%%%%%%%%%%%%%%%%%%%%
%%%%%%%%%%%%%%%%%%%%%%%%%%% begin LEMMA 2 %%%%%%%%%%%%%%%%%%%%%%%%%%%%%%%%%%%%%%%%%%%%%%%%%%%%%%%%%%%%%%%%%%%%%%%%%%%%%%%%%%%%%%%%%%%%%%%%%%%%%%%%%%%%%%%%%%%%%%%%%%
%%%%%%%%%%%%%%%%%%%%%%%%%%% begin LEMMA 2 %%%%%%%%%%%%%%%%%%%%%%%%%%%%%%%%%%%%%%%%%%%%%%%%%%%%%%%%%%%%%%%%%%%%%%%%%%%%%%%%%%%%%%%%%%%%%%%%%%%%%%%%%%%%%%%%%%%%%%%%%%
{\bf Lemma 2.} {\it The PDF and CDF of end-to-end SNR of G2A link are obtained respectively as
\begin{align}
\label{gu1}
f_{\gamma_{\textrm{gu}}}(\gamma_{\textrm{gu}})&=   
\sum_{d_{r}=0}^{jD-1} 
\frac{J_{d_{r}}(\theta'_{r,xy},\sigma^2_{ro})        (J'_{d_{r}})^m       }
{\Gamma(m) (P_t h_L(Z))^m} \nonumber \\
%--------------------- 
&~~~~~~~~~~~~\times \gamma_{\textrm{gu}}^{m-1} 
\exp\left(-\frac{J'_{d_{r}}  \gamma_{\textrm{gu}}  } {P_t h_L(Z)}  \right), 
\end{align}
and
\begin{align}
\label{cug}
F_{\gamma_{\textrm{gu}}}(\gamma_{\textrm{gu}}) &=
\sum_{d_{r}=0}^{jD-1} 
\frac{J_{d_{r}}(\theta'_{r,xy},\sigma^2_{ro})           }
{\Gamma(m)} 
\times\mathcal{V}\left(m, \frac{J'_{d_{r}} \, \gamma_{\textrm{gu}} }{P_t h_L(Z)} \right),
\end{align}
where 
\begin{align}
J'_{d_{r}}\! =\!\left \{\!\!\!
%%%%%%%%%%%%%%%%%%%%%%%%
\begin{array}{ll}
  \frac{m\,\sigma^2}
{2 k^2 d_a^2  G_{\textrm{t,max}} G_0''(N_{r})}, 
~~~&d_{r}=0,\\ \\
%%------------------------------------
   \frac{m\,\sigma^2}
{ G_{t,\textrm{max}} G_0''(N_{r}) \frac{D^2\left(1-\cos\left( \frac{d_{r} k d_a}{D}\right)\right)}{d_{r}^2}}, 
\!\!&d_{r}\in\{0,...,N_{r}\}.
\end{array}
%%%%%%%%%%%%%%%%%%%%%%%%
\right. \nonumber
\end{align}}

%%%%%%%%%%%%%%%%%%%%%%%%%%% END LEMMA 2 %%%%%%%%%%%%%%%%%%%%%%%%%%%%%%%%%%%%%%%%%%%%%%%%%%%%%%%%%%%%%%%%%%%%%%%%%%%%%%%%%%%%%%%%%%%%%%%%%%%%%%%%%%%%%%%%%%%%%%%%%%
%%%%%%%%%%%%%%%%%%%%%%%%%%% END LEMMA 2 %%%%%%%%%%%%%%%%%%%%%%%%%%%%%%%%%%%%%%%%%%%%%%%%%%%%%%%%%%%%%%%%%%%%%%%%%%%%%%%%%%%%%%%%%%%%%%%%%%%%%%%%%%%%%%%%%%%%%%%%%%
%%%%%%%%%%%%%%%%%%%%%%%%%%% END LEMMA 2 %%%%%%%%%%%%%%%%%%%%%%%%%%%%%%%%%%%%%%%%%%%%%%%%%%%%%%%%%%%%%%%%%%%%%%%%%%%%%%%%%%%%%%%%%%%%%%%%%%%%%%%%%%%%%%%%%%%%%%%%%%

%%%%%%%%%%%%%%%%%%%%%%%%%%% begin PROOF %%%%%%%%%%%%%%%%%%%%%%%%%%%%%%%%%%%%%%%%%%%%%%%%%%%%%%%%%%%%%%%%%%%%%%%%%%%%%%%%%%%%%%%%%%%%%%%%%%%%%%%%%%%%%%%%%%%%%%%%%%
%%%%%%%%%%%%%%%%%%%%%%%%%%% begin PROOF %%%%%%%%%%%%%%%%%%%%%%%%%%%%%%%%%%%%%%%%%%%%%%%%%%%%%%%%%%%%%%%%%%%%%%%%%%%%%%%%%%%%%%%%%%%%%%%%%%%%%%%%%%%%%%%%%%%%%%%%%%
%%%%%%%%%%%%%%%%%%%%%%%%%%% begin PROOF %%%%%%%%%%%%%%%%%%%%%%%%%%%%%%%%%%%%%%%%%%%%%%%%%%%%%%%%%%%%%%%%%%%%%%%%%%%%%%%%%%%%%%%%%%%%%%%%%%%%%%%%%%%%%%%%%%%%%%%%%%
%%%%%%%%%%%%%%%%%%%%%%%%%%% begin PROOF %%%%%%%%%%%%%%%%%%%%%%%%%%%%%%%%%%%%%%%%%%%%%%%%%%%%%%%%%%%%%%%%%%%%%%%%%%%%%%%%%%%%%%%%%%%%%%%%%%%%%%%%%%%%%%%%%%%%%%%%%%
%%%%%%%%%%%%%%%%%%%%%%%%%%% begin PROOF %%%%%%%%%%%%%%%%%%%%%%%%%%%%%%%%%%%%%%%%%%%%%%%%%%%%%%%%%%%%%%%%%%%%%%%%%%%%%%%%%%%%%%%%%%%%%%%%%%%%%%%%%%%%%%%%%%%%%%%%%%
\begin{IEEEproof}
In practice, the orientation fluctuations of a fixed ground node is much smaller than the UAV node. Hence, for a ground node, we assume $\theta_{q}\simeq0$. Under such conditions, it is assumed that the fixed ground antenna is perfectly aligned to the antenna mounted on UAV. Hence, the ground antenna gain can be well approximated by its maximum gain at the main-lobe. From this, for G2A link, \eqref{pk1} can be simplified as
\begin{align}
\label{pk2}
\mathbb{G}_{\textrm{gu}}(\theta_{{rx}},\theta_{{ry}}) 
= G_{{t,max}} \times G_r(\theta_{{rx}},\theta_{{ry}}),
\end{align}
%and
%\begin{align}
%\label{pk3}
%\mathbb{G}_{\textrm{ug}}(\theta_{\textrm{tx}},\theta_{\textrm{ty}}) 
%=  G_{\textrm{r,max}} \times G_t(\theta_{\textrm{tx}},\theta_{\textrm{ty}}).
%\end{align}
%where $G_{\textrm{t,max}}= G_t(\theta_{\textrm{tx}}\simeq0,\theta_{\textrm{ty}}\simeq0)$ and 
%$G_{\textrm{r,max}}= G_r(\theta_{\textrm{rx}}\simeq0,\theta_{\textrm{ry}}\simeq0)$.
where $G_{t,\textrm{max}}= G_t(\theta_{{tx}}\simeq0,\theta_{{ty}}\simeq0)$.
From \eqref{pk2} and similar to the derivation of \eqref{ro}, we have
\begin{align}
\label{rt1}
&f_{\mathbb{G}_{\textrm{gu}}}(\mathbb{G}_{\textrm{gu}})=   
J_0(\theta'_{r,xy},\sigma^2_{ro})\,\delta\left(G_{r}-2 k^2 d_a^2 G_{t,\textrm{max}} G_0''(N_{r}) \right)\\
%---------------------------
&~~~~~~~~~+\sum_{d_{r}=1}^{jD-1} J_{d_{r}}(\theta'_{r,xy},\sigma^2_{ro})\nonumber\\
%--------------------------- 
&~~~~~~~~~\times\delta\left(G_{r}- 
G_{t,\textrm{max}} G_0''(N_{r}) \frac{D^2\left(1-\cos\left( \frac{d_{r} k d_a}{D}\right)\right)} {d_{r}^2}\right). \nonumber
\end{align}
Finally, using \eqref{ss1}, \eqref{Gamma} and \eqref{rt1}, the closed-form expression of the G2A channel PDF is derived in \eqref{gu1}.
\end{IEEEproof}
%%%%%%%%%%%%%%%%%%%%%%%%%%% END PROOF %%%%%%%%%%%%%%%%%%%%%%%%%%%%%%%%%%%%%%%%%%%%%%%%%%%%%%%%%%%%%%%%%%%%%%%%%%%%%%%%%%%%%%%%%%%%%%%%%%%%%%%%%%%%%%%%%%%%%%%%%%%%
%%%%%%%%%%%%%%%%%%%%%%%%%%% END PROOF %%%%%%%%%%%%%%%%%%%%%%%%%%%%%%%%%%%%%%%%%%%%%%%%%%%%%%%%%%%%%%%%%%%%%%%%%%%%%%%%%%%%%%%%%%%%%%%%%%%%%%%%%%%%%%%%%%%%%%%%%%%%
%%%%%%%%%%%%%%%%%%%%%%%%%%% END PROOF %%%%%%%%%%%%%%%%%%%%%%%%%%%%%%%%%%%%%%%%%%%%%%%%%%%%%%%%%%%%%%%%%%%%%%%%%%%%%%%%%%%%%%%%%%%%%%%%%%%%%%%%%%%%%%%%%%%%%%%%%%%%
%%%%%%%%%%%%%%%%%%%%%%%%%%% END PROOF %%%%%%%%%%%%%%%%%%%%%%%%%%%%%%%%%%%%%%%%%%%%%%%%%%%%%%%%%%%%%%%%%%%%%%%%%%%%%%%%%%%%%%%%%%%%%%%%%%%%%%%%%%%%%%%%%%%%%%%%%%%%
%%%%%%%%%%%%%%%%%%%%%%%%%%% END PROOF %%%%%%%%%%%%%%%%%%%%%%%%%%%%%%%%%%%%%%%%%%%%%%%%%%%%%%%%%%%%%%%%%%%%%%%%%%%%%%%%%%%%%%%%%%%%%%%%%%%%%%%%%%%%%%%%%%%%%%%%%%%%
%%%%%%%%%%%%%%%%%%%%%%%%%%% END PROOF %%%%%%%%%%%%%%%%%%%%%%%%%%%%%%%%%%%%%%%%%%%%%%%%%%%%%%%%%%%%%%%%%%%%%%%%%%%%%%%%%%%%%%%%%%%%%%%%%%%%%%%%%%%%%%%%%%%%%%%%%%%%
%%%%%%%%%%%%%%%%%%%%%%%%%%% END PROOF %%%%%%%%%%%%%%%%%%%%%%%%%%%%%%%%%%%%%%%%%%%%%%%%%%%%%%%%%%%%%%%%%%%%%%%%%%%%%%%%%%%%%%%%%%%%%%%%%%%%%%%%%%%%%%%%%%%%%%%%%%%%
\textcolor{white}{p}
\textcolor{white}{s}

As one can observe, the proposed channel model in \eqref{gu1} is a simple function of end-to-end SNR $\gamma_\textrm{gu}$. However, despite the simplicity and tractability, the proposed closed-form channel model composes the effects of large- and small-scale fading characterized by $\alpha$, $m$ and $h_L(Z)$, UAVs stability characterized by $\theta'_{rx}$, $\theta'_{ry}$, and $\sigma^2_{ro}$, and antenna pattern specifications characterized by $N_t$ and $N_r$.

The channel distribution of the A2G link can be obtained similarly from \eqref{gu1} by swapping subscript $t$ with subscript $r$ and vice versa.

%
%%%%%%%%%%%%%%%%%%%%%%%%%%%%%%%%%%%%%%%%%%%%%%%%%%%%%%%%%%%%%%%%
%%%%%%%%%%%%%%%%%%%%%%%%%%%%%%%%%%%%%%%%%%%%%%%%%%%%%%%%%%%%%%%% VERSUS W_Z
\begin{figure}
	\centering
	\subfloat[] {\includegraphics[width=2.75 in]{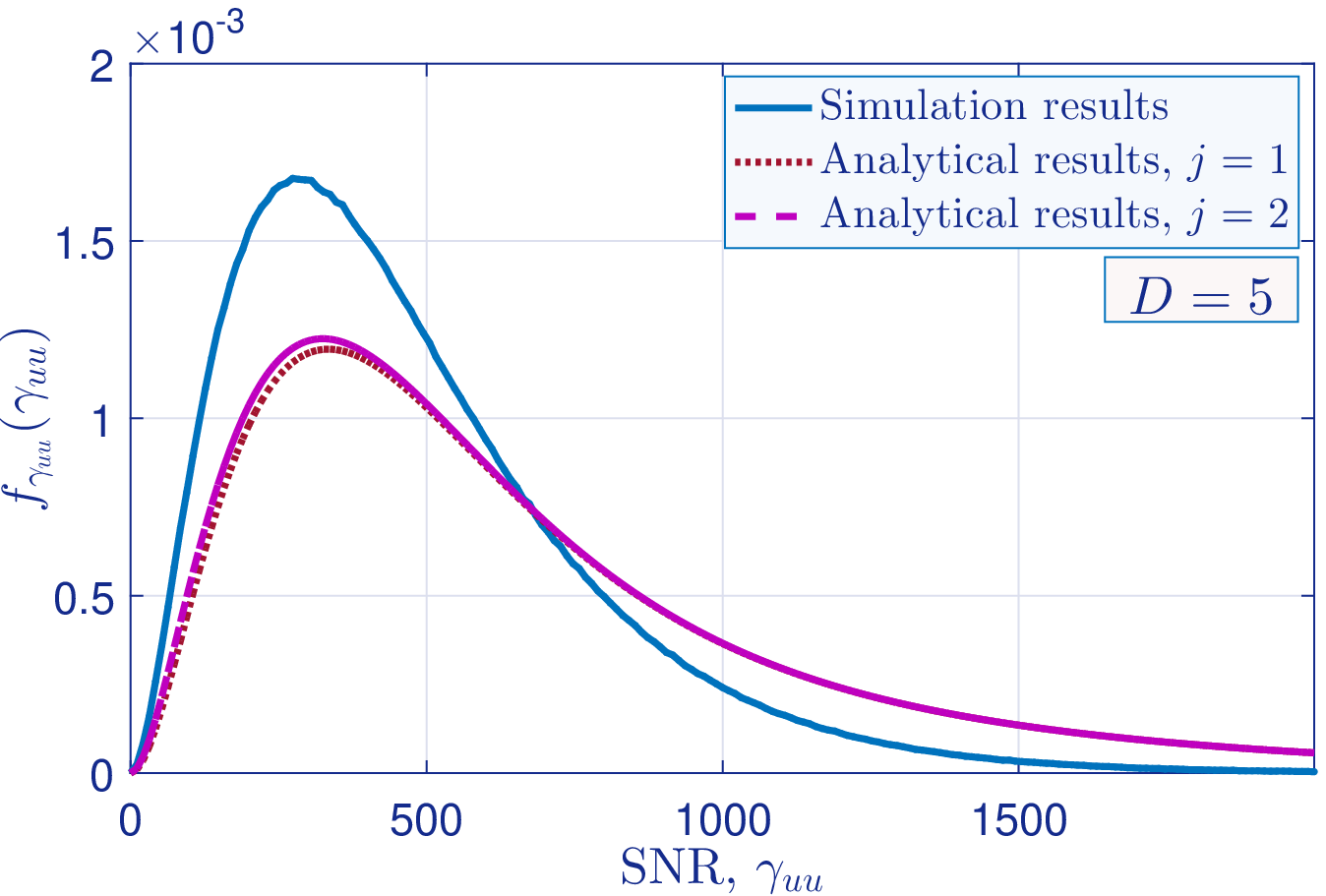}
		\label{xs1}
	}
	\hfill
	\subfloat[] {\includegraphics[width=2.75 in]{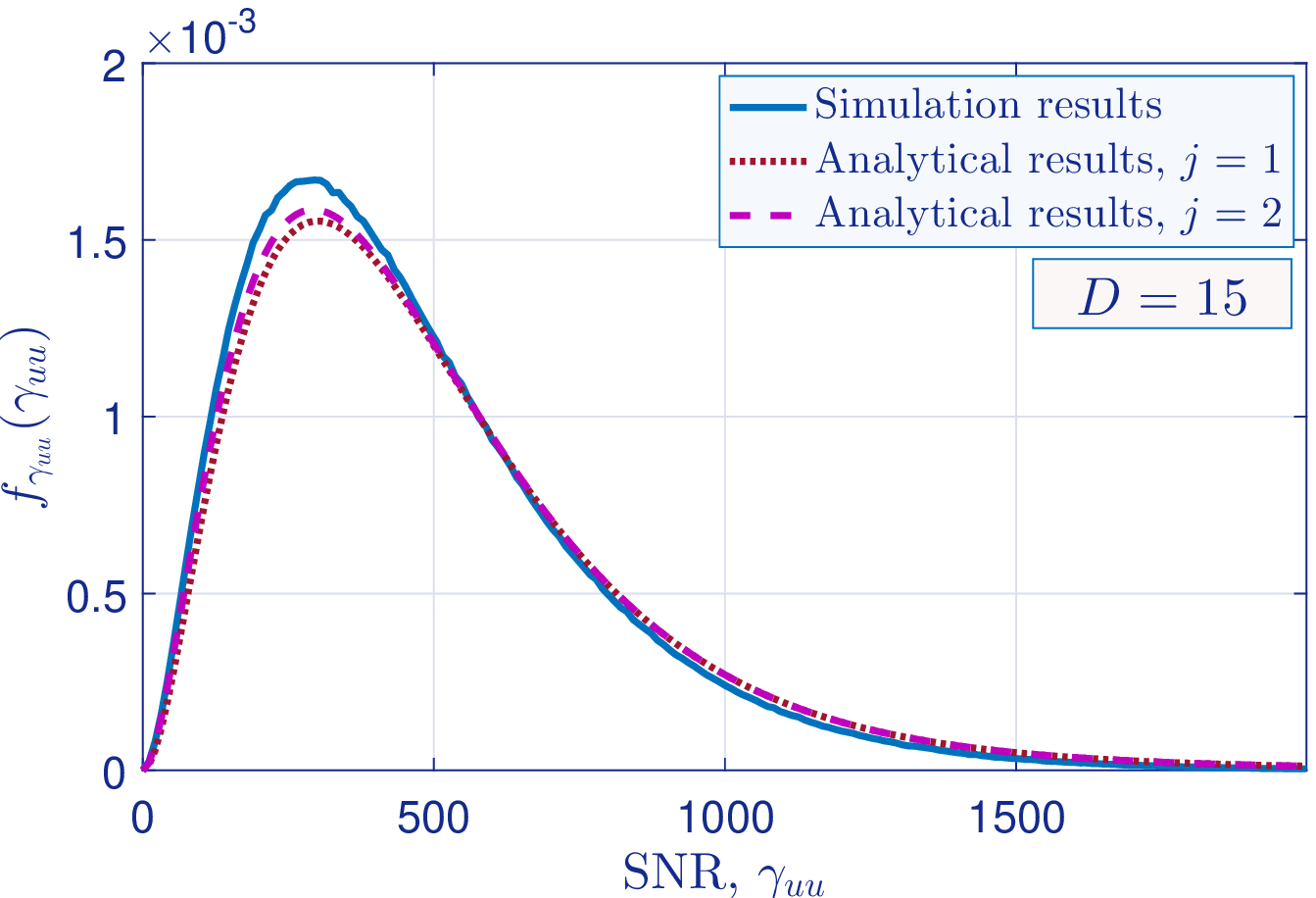}
		\label{xs2}
	}
	\hfill
	\subfloat[] {\includegraphics[width=2.75 in]{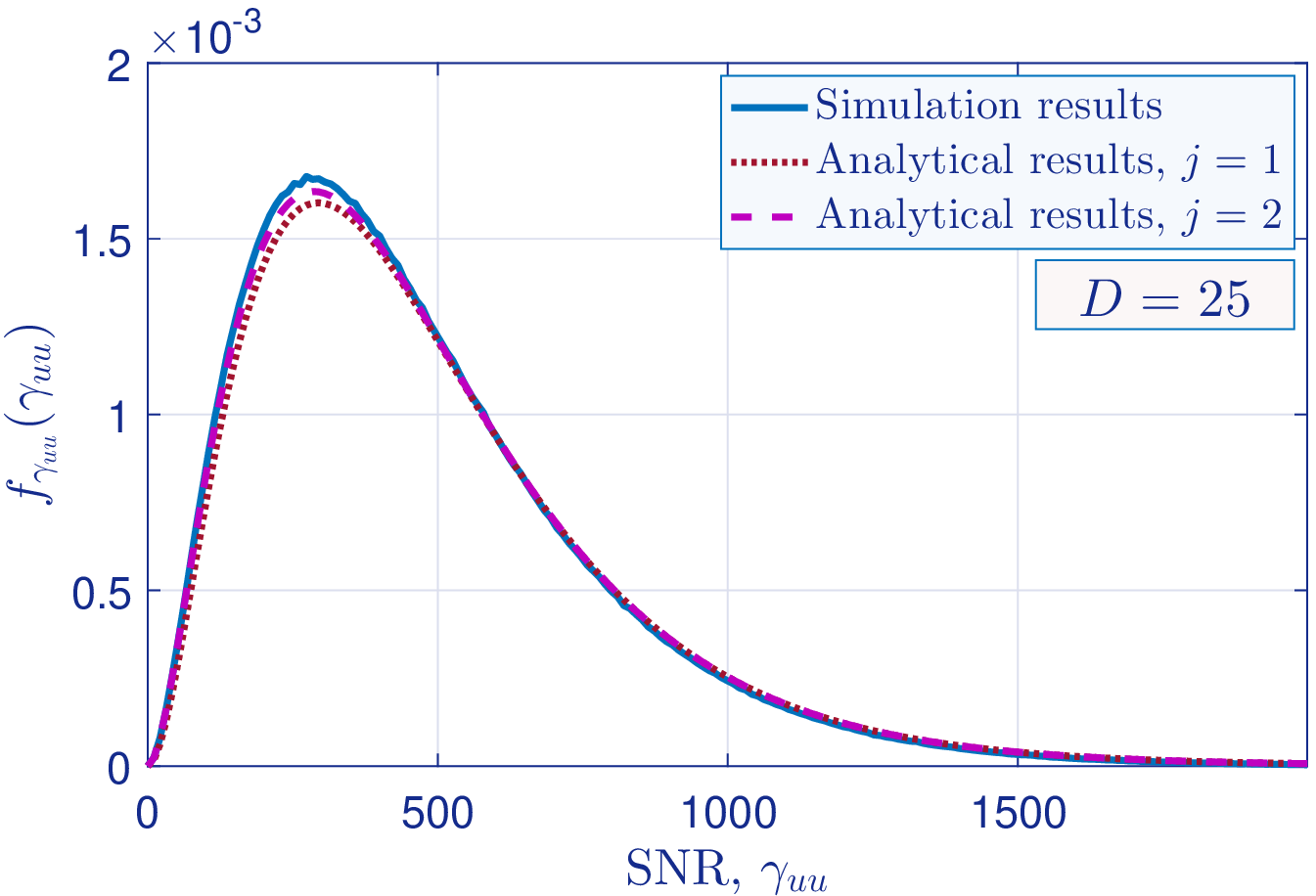}
		\label{xs3}
	}
	\caption{Channel distribution of A2A link when $\sigma^2_{to}=\sigma^2_{ro}=1^o$, $\theta'_{tx}=\theta'_{ty}=0.5^o$ and $\theta'_{rx}=\theta'_{ry}=1^0$,  for a) $D=5$, b) $D=15$, and c) $D=25$.}
	\label{xss}
\end{figure}
%%%%%%%%%%%%%%%%%%%%%%%%%%%%%%%%%%%%%%%%%%%%%%%%%%%%%%%%%%%%%%%%
%%%%%%%%%%%%%%%%%%%%%%%%%%%%%%%%%%%%%%%%%%%%%%%%%%%%%%%%%%%%%%%%
%
%
%%%------------------------------------------
%%%------------------------------------------
\section{Simulation Results}
%%%------------------------------------------
%%%------------------------------------------
%For performance evaluation, we perform Monte Carlo simulations with $5\times10^{7}$ independent s.

%
%%%%%%%%%%%%%%%%%%%%%%%%%%%%%%%%%%%%%%%%%%%%%%%%%%%%%%%%%%%%%%%%
%%%%%%%%%%%%%%%%%%%%%%%%%%%%%%%%%%%%%%%%%%%%%%%%%%%%%%%%%%%%%%%% VERSUS W_Z
\begin{figure}
	\centering
	\subfloat[] {\includegraphics[width=2.8 in]{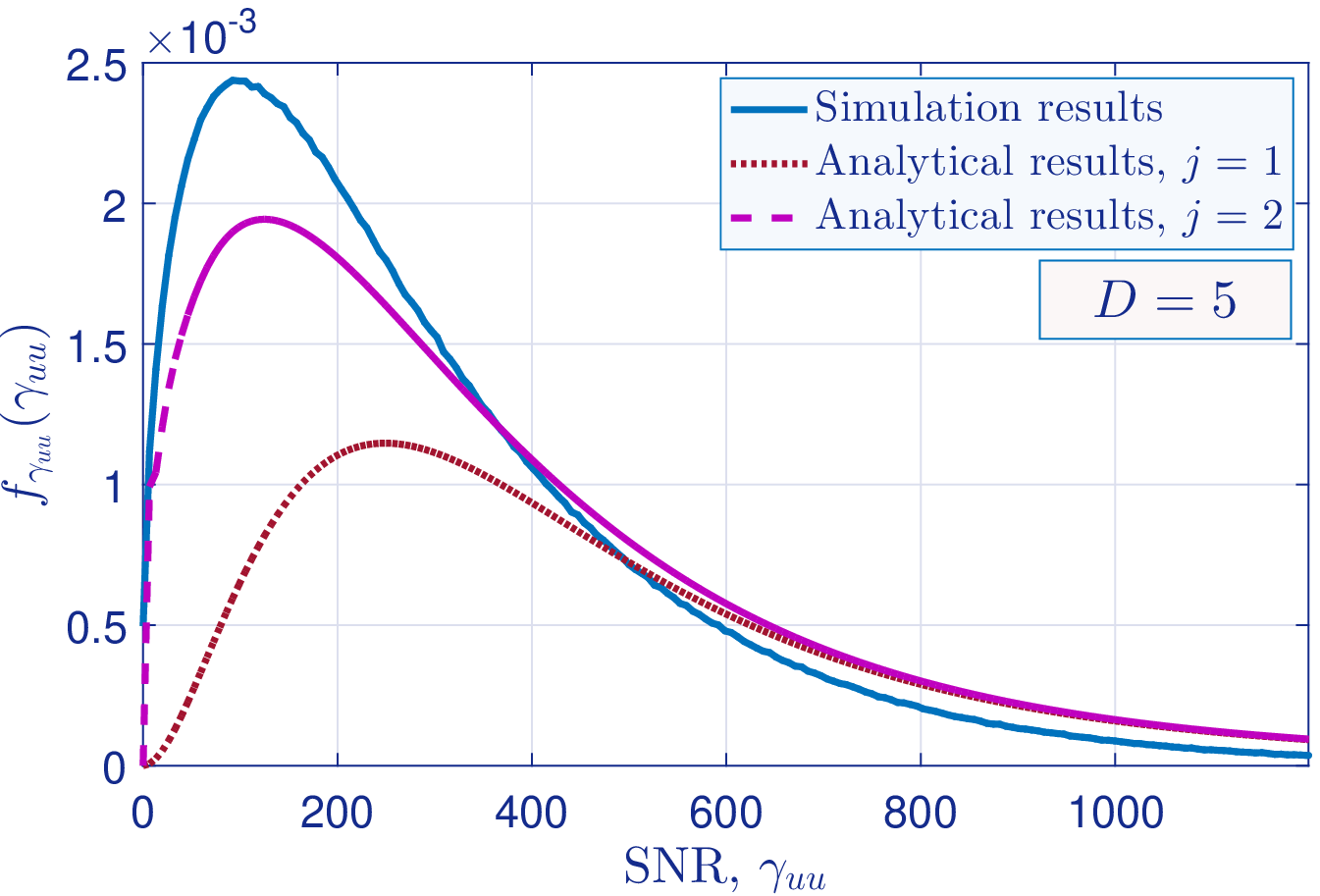}
		\label{cs1}
	}
	\hfill
	\subfloat[] {\includegraphics[width=2.8 in]{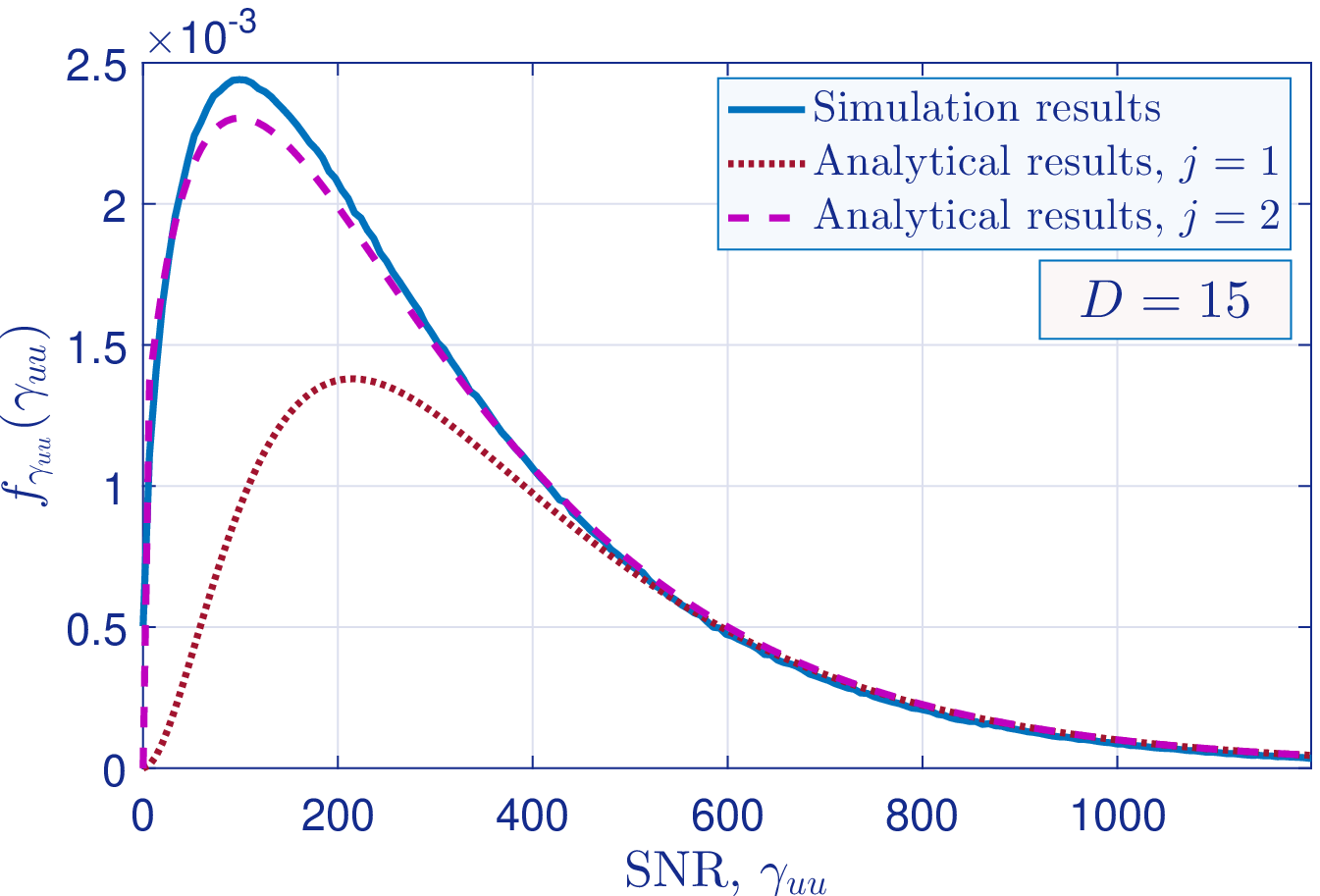}
		\label{cs2}
	}
	\hfill
	\subfloat[] {\includegraphics[width=2.8 in]{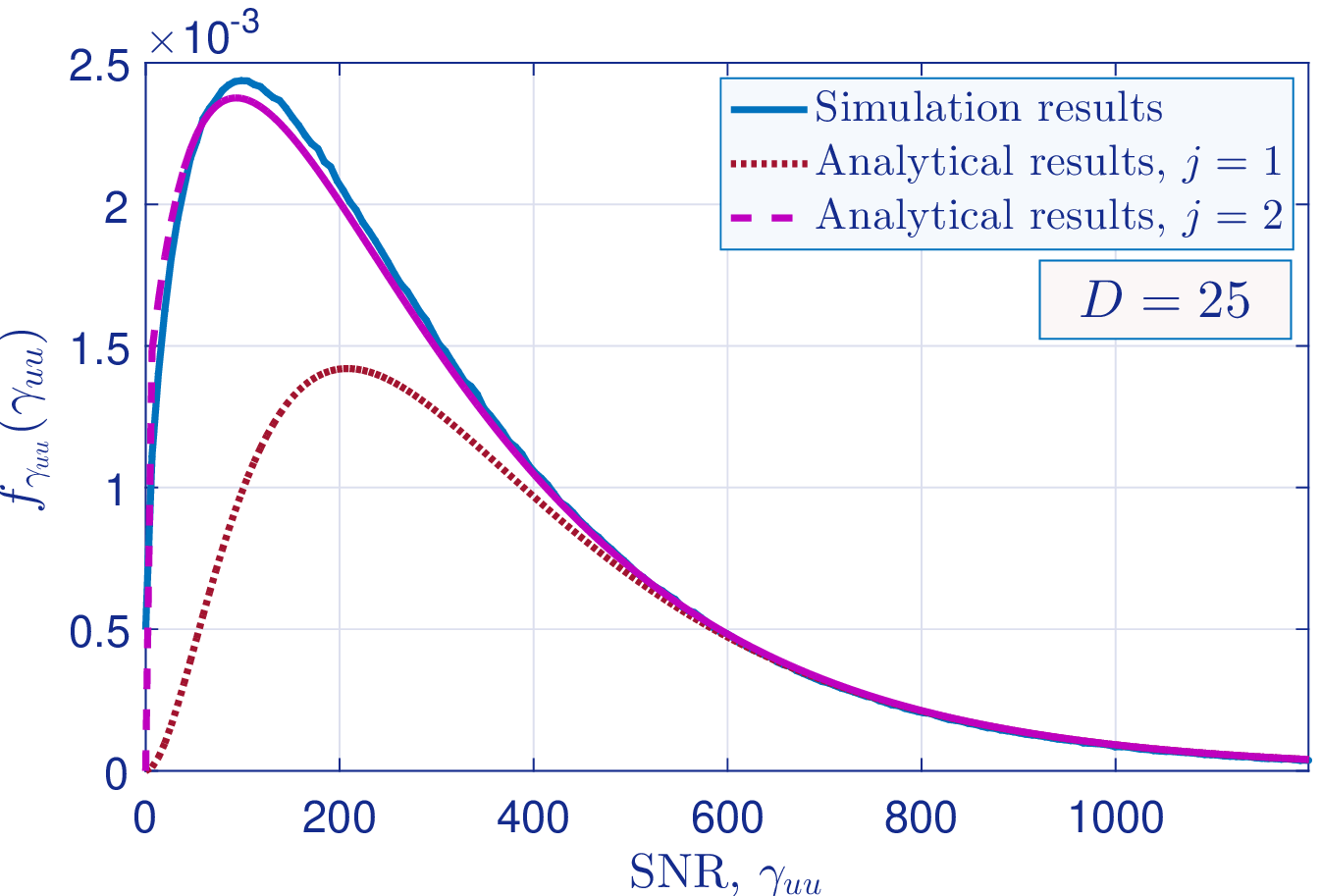}
		\label{cs3}
	}
	\caption{Channel distribution of A2A link when $\sigma^2_{to}=\sigma^2_{ro}=3^o$, $\theta'_{tx}=\theta'_{ty}=1^o$ and $\theta'_{rx}=\theta'_{ry}=0.5^o$,  for a) $D=5$, b) $D=15$, and c) $D=25$.}
	\label{css}
\end{figure}
%%%%%%%%%%%%%%%%%%%%%%%%%%%%%%%%%%%%%%%%%%%%%%%%%%%%%%%%%%%%%%%%
%%%%%%%%%%%%%%%%%%%%%%%%%%%%%%%%%%%%%%%%%%%%%%%%%%%%%%%%%%%%%%%%
%

%For our simulations, we consider the UAVs to have the same standard deviation of AoA and AoD fluctuations, i.e., for the U2U link 
For our simulations, we consider standard values for system parameters, as follows.
The carrier frequency $f_c=50$~GHz, average building height $h_b=30$~m, thermal noise power $\sigma^2=-110$~dBm, Nakagami fading parameter $m=3$, transmit power $P_t=20$~dBm, and SNR threshold $\gamma_{\textrm{th}}=10$~dB.
%%%%%%%%%%%%%%%%%%%%%%%%%%%%%%%%%%%%%%%%%%%%%%%%%%%%%%%%%%%%%%%%%%%
\subsubsection{Accuracy of the Derived Channel Models}
%%%%%%%%%%%%%%%%%%%%%%%%%%%%%%%%%%%%%%%%%%%%%%%%%%%%%%%%%%%%%%%%%%%
The accuracy of the proposed channel PDF for a A2A link is evaluated in Figs. \ref{xss} and \ref{css} for two different values of the UAV's instability parameters $\sigma_{to}=\sigma_{ro}=1^o$ and $\sigma_{to}=\sigma_{ro}=3^o$, respectively.
The accuracy of the analytical channel model is validated using Monte Carlo simulations. For Monte Carlo simulations, we generate $5\times10^{7}$ independent RVs $\alpha$, $\theta_{{tx}}$, $\theta_{{ty}}$, $\theta_{{rx}}$, and $\theta_{{ry}}$. For each $5\times10^{7}$ independent run, we calculate antenna pattern from actual model given in \eqref{p_1}-\eqref{pk1}, and then, calculate $5\times10^{7}$ instantaneous SNR from \eqref{ss1}. 
%Finally, by using {\it hist()} command of MATLAB software, the actual channel model is plotted to test the accuracy of proposed analytical model.
%
From \eqref{uu}, $D$ and $j\in\{1,2\}$ are two parameters that impact on the validity of channel PDF.
The variable $D$ is used for approximating the antenna pattern. The optimal value of $D$ is its minimum value that satisfies a predefined accuracy. Parameter $j\in\{1,2\}$ is another parameter that determines the accuracy of the derived channel PDF. 
%The value $j=1$ is used when the main-lobe of pattern is considered and for more precision, $j=2$ is used when main-lobe along with the first side-lobe is considered. 
The results of Figs. \ref{xss} and \ref{css} are plotted for 
%UAVs with angular stability $\sigma_\textrm{to}=\sigma_\textrm{ro}=1^o$ and 
different values of $D$ and two different values of $j=1$ and $j=2$.
Figs. \ref{xss} and \ref{css} demonstrate that $D=25$ can be a good choice. Moreover, by comparing the results of these figure, it can be observed that for UAVs with higher angular stability (lower $\sigma_{to}$ and $\sigma_{ro}$), $j=1$ is a good choice. Meanwhile, by decreasing stability (increasing $\sigma_{to}$ and $\sigma_{ro}$), the accuracy of analytical channel model decreases for $j=1$. Under this condition, the channel must be modeled with main-lobe along with the first side-lobe, i.e., $j=2$.

%%%%%%%%%%%%%%%%%%%%%%%%%%%%%
%%%%%%%%%%%%%%%%%%%%%%%%%%%%%
\subsubsection{Performance Analysis and Optimal Pattern Selection}
%%%%%%%%%%%%%%%%%%%%%%%%%%%%%
%%%%%%%%%%%%%%%%%%%%%%%%%%%%%

\begin{figure}
	\begin{center}
		\includegraphics[width=3.3 in ]{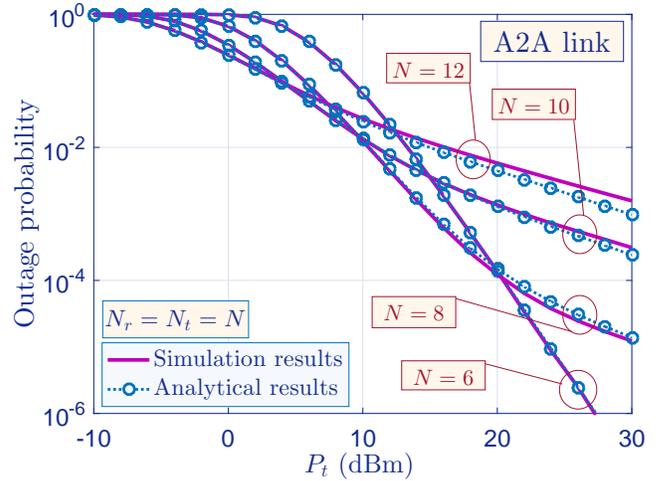}
		\caption{Outage probability of A2A link for $\sigma_{to}=\sigma_{ro}=2^o$, $\theta'_{tx}=\theta'_{ty}=\theta'_{rx}=\theta'_{ry}=0.5^o$, and different values of $N_t$ and $N_r$.}
		\label{zx1}
	\end{center}
\end{figure}

Next, the performance of the considered UAV-based mmW link is studied in terms of outage probability.
To study the impact of antenna pattern on the system performance, in Fig \ref{zx1}, the outage probability of A2A link is plotted versus $P_t$ and for different values of $N_t=N_r=N$.
As we observe, the accuracy of the derived closed-form expression for the outage probability is verified via our simulation results.   
%Due to the possibility of transmitting with higher antenna gain, at low values of SNR, the more numbers of antennas, the better performance of the system.
From this figure, we can see that lower values for $N$ achieve a better performance at a high $P_t$ regime. However, at a low $P_t$ regime, higher values of $N$ achieve better performance. This can be justified since the poor SNR at low values of $P_t$ can be compensated by high directional antenna pattern. Meanwhile, at high values of $P_t$, the outage probability of high directional beam is floored due to the UAVs' orientation fluctuations. Under such conditions, antenna pattern must be selected wider to compensate UAVs' orientation fluctuations. 

\begin{figure}
	\begin{center}
		\includegraphics[width=3.4 in ]{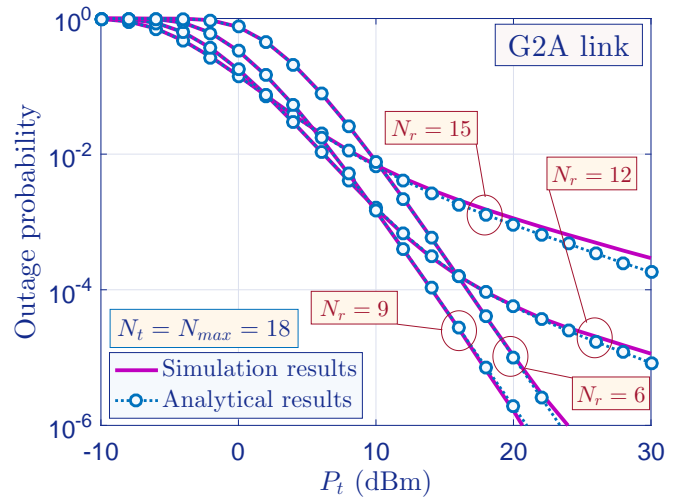}
		\caption{Outage probability of G2A link for $\sigma_{ro}=2^o$, $\theta'_{rx}=\theta'_{ry}=0.5^o$, $N_t=N_\textrm{max}=18$, and different values of $N_r$.}
		\label{zx2}
	\end{center}
\end{figure}

In Fig. \ref{zx2}, we investigate the impact of antenna pattern on the outage probability of G2A link. As mentioned, for a G2A link, due to the high stability of a fixed ground antenna compared to the aerial antenna, we consider $N_t=N_\textrm{max}=18$. The results of Fig. \ref{zx2} are obtained for $\sigma_{ro}=2^o$, $\theta'_{rx}=\theta'_{ry}=0.5^o$ and $Z=2000$ m. 
Similar to the A2A link, we can see that higher values of $N_r$ achieve a better performance at a low $P_t$ regime, and vice versa.
Meanwhile, we observe a perfect match between the analytical and simulation-based results which validates the accuracy of our derived analytical expressions for G2A link.
By comparing the results of Figs. \ref{zx1} and \ref{zx2}, as expected, we can observe that the G2A link achieves  better performance compared to the A2A link. 

\begin{figure}
	\begin{center}
		\includegraphics[width=3.3 in ]{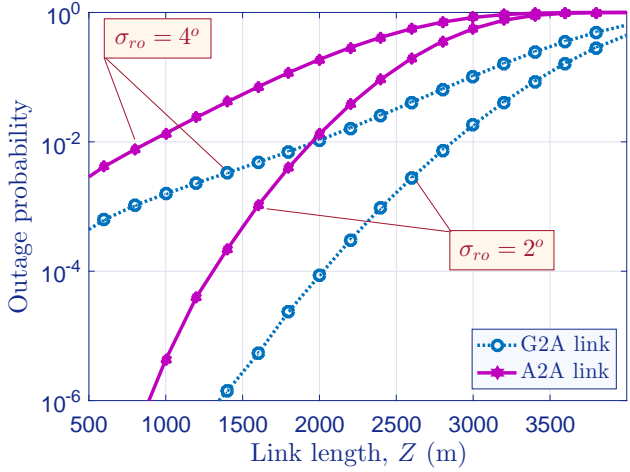}
		\caption{Comparison of the outage probability of A2A and G2A links versus link length for two different conditions, $\sigma_{ro}=2^o$ and $\sigma_{ro}=4^o$. For G2A link, $N_t=N_\textrm{max}$ and $N_r=8$, and for A2A link, $\sigma_{to}=\sigma_{ro}$ and $N_t=N_r=8$.}
		\label{zx3}
	\end{center}
\end{figure}

To have a better comparing between A2A and G2A links, we evaluated the outage probability of both links in Fig. \ref{zx3} as a function of the link length, $Z$. The results of Fig. \ref{zx3} are provided for $P_t=14$ dBm and two different values for the UAV instability factor $\sigma_{ro}=2^o$ and $\sigma_{ro}=4^o$. As expected, by increasing link length, the channel loss increases, and thus, performance degrades. However, increasing link length has a more sever effect on the A2A link compared to the G2A link. Fig. \ref{zx3} shows that, for a link length greater than 2000\,m, a G2A link with $\sigma_{ro}=4^o$ achieves a lower outage probability than an A2A link with $\sigma_{ro}=2^o$.
The results of Fig. \ref{zx3} are provided for the constant values for $N_t$ and $N_r$. However, we expect that by varying link length, the optimal value for $N_t$ and $N_r$ change.

\begin{figure}
	\begin{center}
		\includegraphics[width=3.3 in ]{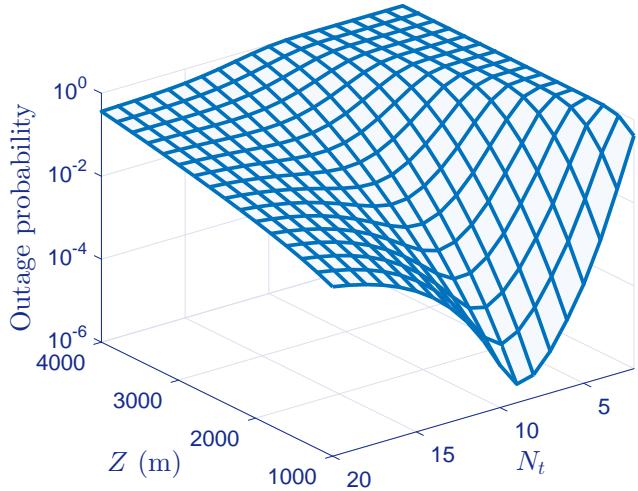}
		\caption{Outage probability of A2A link versus link length and $N_t$ for $N_r=N_t$, $\sigma_{to}=\sigma_{ro}=2$ mrad, and $P_t=15$ dBm.}
		\label{zx4}
	\end{center}
\end{figure}

In Fig. \ref{zx4}, we investigate the outage probability of A2A link versus $Z$ and $N$ in order to shed light on the impact of the link length on the optimal value of $N$. This figure clearly shows that by varying link length, the optimal value for $N$ changes. As the link length increases, the optimal value for $N$ must be increased to compensate for the additional channel loss due to the larger link length. For instance, from the results of Fig. \ref{zx4}, we observe that, by increasing the link length from 1000 to 3000 m, the optimal value for $N$ increases from 9 to 15.

\begin{figure}
	\begin{center}
		\includegraphics[width=3.3 in ]{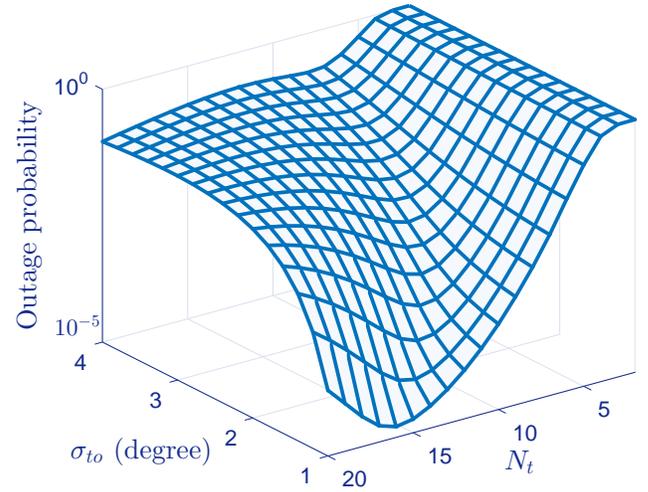}
		\caption{Outage probability of A2A link versus $\sigma_{to}$ and $N_t$ when $\sigma_{ro}=\sigma_{to}$ and $N_r=N_t=N$.}
		\label{zx5}
	\end{center}
\end{figure}

The strength of UAV's orientation fluctuations is the another important parameter that can affect on the optimal value for antenna pattern, $N$. In Fig. \ref{zx5}, the outage probability of A2A link is plotted versus $N_t$ and $\sigma_{to}$ for a special symmetrical case wherein $\sigma_{to}=\sigma_{ro}$ and $N_t=N_r=N$. From Fig. \ref{zx5}, we can see that, for the symmetrical case, the optimal value for $N_t$ decreases by increasing $\sigma_{to}$. For instance, by increasing $\sigma_{to}=1^o$ to $\sigma_{to}=3^o$, the optimal value for $N_t$ decreases from 16 to 9. This can be justified since by increasing $\sigma_{to}$, the beamwidth of antenna pattern must be increases to compensate the orientation fluctuations of the hovering UAV. Hence, the antenna gain must be decreased in order to increase antenna beamwidth. The results of Fig. \ref{zx5} are provided for a symmetrical  case. 

%%%%%%%%%%%%%%%%

\begin{figure}
	\centering
	\subfloat[] {\includegraphics[width=3.3 in]{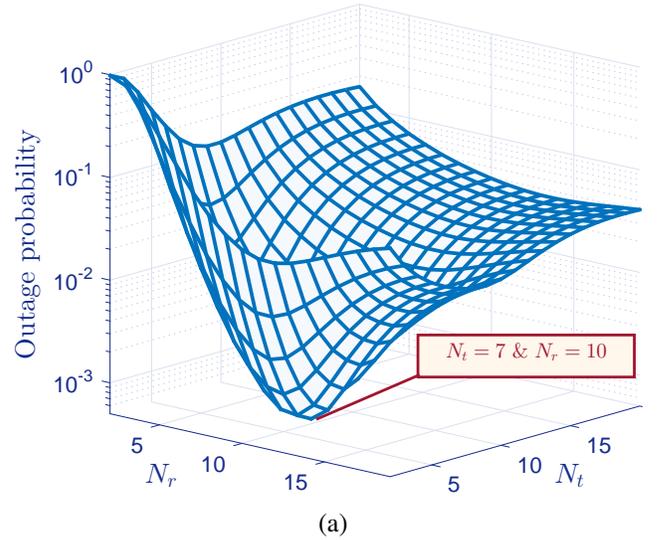}
		\label{cx1}
	}
	\hfill
	\subfloat[] {\includegraphics[width=3.3 in]{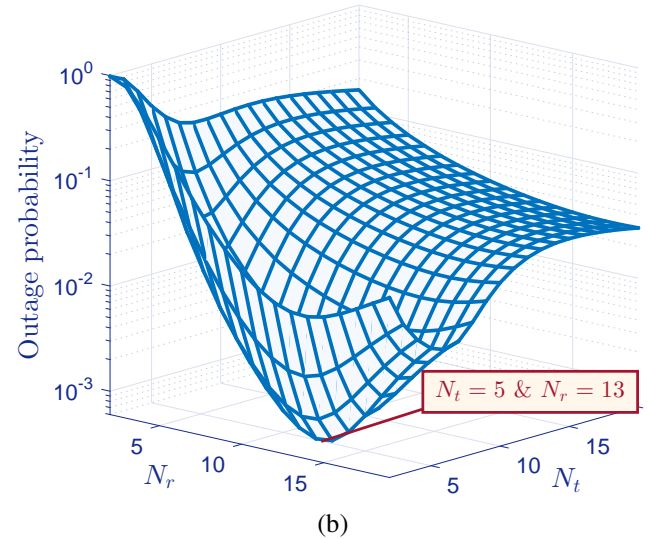}
		\label{cx2}
	}
	\caption{Outage probability of A2A link versus $N_{t}$ and $N_{r}$ for two different scenario with different instabilities: 
		(a) for $\sigma_{to}=3^o$ and $\sigma_{ro}=2^o$, and (b) for $\sigma_{to}=4^o$ and $\sigma_{ro}=1.5^o$.}
	\label{cxs}
\end{figure}

To get a better insight about a more general case, in Figs. \ref{cx1} and \ref{cx2} we investigate the outage probability of an A2A link versus $N_t$ and $N_r$ for two different UAVs' angular stability where $\sigma_{to}\neq\sigma_{ro}$.
From these figures, we observe that by changing $N_t$ and $N_r$, the outage probability changes, significantly. More importantly, by changing the angular stability, the optimal antenna pattern changes. For instance, by changing UAVs' angular stability from $\sigma_{to}=3^o, \sigma_{ro}=2^o$ to $\sigma_{to}=4^o, \sigma_{ro}=1.5^o$, the optimal values for $N_t$, $N_r$ changes from $N_t=7, N_r=10$ to $N_t=5, N_r=13$. Here, we note that, in addition to the mechanical control system of UAV, air pressure and wind speed can affect on the UAV's angular stability. Since air pressure and wind speed continuously changes in the day time, it is reasonable to expect that the UAV's angular stability changes in day time. 
Hence, to achieve a reliable communication link for the considered UAV-based system, we propose that to design the square array antenna  with maximum number of antenna elements, $N_\textrm{max}\times N_\textrm{max}$. Then, for any given angular stability, which  continuously changes in the order of several minutes to several hours, we only activate $N_t\times N_t$ Tx antenna elements and $N_r\times N_r$ Rx antenna elements  of  $N_\textrm{max}\times N_\textrm{max}$ that achieves a minimum outage probability where $N_t, N_r\in\{1,...,N_\textrm{max}\}$. Formally, the problem can be formulated as
\begin{align}
\label{dc1}
&\mathop {{\textrm{ min}}}
\limits_{ N_{rs}, N_{rd}, \psi_s, \psi_d} 
~~\mathbb{P}_\textrm{out},\\
%---------------------------------------------
& ~~~~~~
\textrm{ s.t.} ~~~~~~~~
N_{rs}, N_{rd}\in\{1,N_\textrm{max} \} \nonumber.
\end{align}
For different values of $\sigma_{to}$ and $\sigma_{ro}$, the optimal number of $N_t$, $N_r$, and the corresponding minimum achievable outage probabilities are provided in Table \ref{tab1}. The optimal results are obtained by simulations. 
%As we observe, the optimal number of $N_r$ and $N_t$ changes by varying UAVs' angular instability. 
The simulation results are obtained by performing Monte Carlo simulations with $5\times10^{7}$ independent runs and using a computer with an Intel i7-3632QM CPU running at 2.20 GHz with 8 GB RAM. Under such conditions, the running time of the optimization problem takes approximately $500$\,s. 

Finally, to confirm the accuracy of our derived analytical expressions, as a suboptimal method, we numerically  find $N_t$ and $N_r$ by using \eqref{cuu}. The simulation results of Table \ref{tab1} confirm the validity of the numerical results that only requires running time $\simeq 1$. Note that, for $5\times10^{7}$ independent runs, the simulation results is valid when $\mathbb{P}_\textrm{out}>10^6$. For high quality of services with lower outage probability, we require to more independent runs that increases running time. Moreover, in \eqref{dc1}, we only optimize two parameters $N_t$ and $N_r$. 
%
%However, for more optimization parameters such as optimal code rate, modulation size, UAV's aerial position, the processing time exponentially increases which shows the importance of provided analytical closed-form expressions become more.
%
%However, one can use analytical expressions provided in this paper to calculate the amount of outage probability under different conditions in an extremely time efficient manner.
%
However, one can use analytical expressions provided in this paper to optimize the other parameters such as code rate, modulation size, UAV's aerial position and so on, in an extremely time efficient manner.

%%%%%%%%%%%%%%%%%%%%%%%%%%%%%%%%%%%%%%%%%%%%%%%%%%%%%%%%%%%%%%%%
%%%%%%%%%%%%%%%%%%%%%%%%%%%%%%%%%%%%%%%%%%%%%%%%%%%%%%%%%%%%%%%% VERSUS P_T
%%%%%%%%%%%%%%%%%%%%%%%%%%%%%%%%%%%%%%%%%%%%%%%%%%%%%%%%%%%%%%%%

%%%%%%%%%%%%%%%%%%%%%%%%%%%%%%%%%%%%%%%%%%%%%%%%%%%%%%%%%%%%%%%%
%%%%%%%%%%%%%%%%%%%%%%%%%%%%%%%%%%%%%%%%%%%%%%%%%%%%%%%%%%%%%%%% VERSUS P_T
%%%%%%%%%%%%%%%%%%%%%%%%%%%%%%%%%%%%%%%%%%%%%%%%%%%%%%%%%%%%%%%%

%%%%%%%%%%%%%%%%%%%%%%%%%%%%%%%%%%%%%%%%%%%%%%%%%%%%%%%%%%%%%%%%
%%%%%%%%%%%%%%%%%%%%%%%%%%%%%%%%%%%%%%%%%%%%%%%%%%%%%%%%%%%%%%%% VERSUS P_T
%%%%%%%%%%%%%%%%%%%%%%%%%%%%%%%%%%%%%%%%%%%%%%%%%%%%%%%%%%%%%%%%

%%%%%%%%%%%%%%%%%%%%%%%%%%%%%%%%%%%%%%%%%%%%%%%%%%%%%%%%%%%%%%%%
%%%%%%%%%%%%%%%%%%%%%%%%%%%%%%%%%%%%%%%%%%%%%%%%%%%%%%%%%%%%%%%% VERSUS P_T
%%%%%%%%%%%%%%%%%%%%%%%%%%%%%%%%%%%%%%%%%%%%%%%%%%%%%%%%%%%%%%%%

%%%%%%%%%%%%%%%%%%%%%%%%%%%%%%%%%%%%%%%%%%%%%%%%%%%%%%%%%%%%%%%%
%%%%%%%%%%%%%%%%%%%%%%%%%%%%%%%%%%%%%%%%%%%%%%%%%%%%%%%%%%%%%%%% VERSUS P_T
%%%%%%%%%%%%%%%%%%%%%%%%%%%%%%%%%%%%%%%%%%%%%%%%%%%%%%%%%%%%%%%%

%
%%%%%%%%%%%%%%%%%%%%%%%%%%%%%%%%%%%%%%%%%%%%%%%%%%%%%%%%%%%%%%%%
%%%%%%%%%%%%%%%%%%%%%%%%%%%%%%%%%%%%%%%%%%%%%%%%%%%%%%%%%%%%%%%% VERSUS W_Z
\section{Conclusion}
%%---------------------------------------------------------------------------
%%---------------------------------------------------------------------------
%%---------------------------------------------------------------------------
%%%%%%%%%%%%%%%%%%%%%%%%
In this paper, we have studied the performance of UAV-based mmW links when UAVs are equipped with square array antenna.
Accordingly, we have characterized the UAV-based mmW channels by considering the large scale path loss, small scale fading along with antenna patterns as well as the random effect of UAVs' angular vibrations.
For performance analysis, we have derived closed-form statistical channel models for A2A, G2A, and A2G channels. We have then verified the accuracy of analytical models by employing Monte Carlo simulations. 
%
%Then, analytical results have been used to study the effect of antenna pattern gain under different severity of UAVs' angular vibrations for establishing reliable UAV-assisted  mmW links in terms of achieving minimum outage probability.
%
Our analytical results have made it possible to find the optimal antenna directivity gain for designing a reliable UAV-based mmW communications  under different levels of stability of UAVs without resorting to time-consuming simulations.
%
%Analytical results have shown that lower pattern gain achieves better performance at high $P_t$ regime and vise versa.
%
%

%
%
%%%%%%%%%%%%%%%%%%%%%%%%% TABEL of simple UAV system
\begin{table*}
	\def\tablename{Table}
	\centering
	\caption{Comparison of the Optimal values for $N_{\rm t}$ and $N_{\rm r}$ obtained by simulation and numerical results to achieve minimum outage probability over A2A link for different conditions of UAV instabilities. 
		%$\sigma_{\rm to}~\textrm{and}~\sigma_{\rm ro}$	when $P_t=15~{\rm dBm}$.
	}
	% $\theta'_\textrm{tx}=\theta'_\textrm{ty}=\theta'_\textrm{rx}=\theta'_\textrm{ry}=1^o$
	\begin{tabular}{|c|c||   
			c|c|c|c||     
			c|c|c|c|}
		\cline{1-10}
		\multicolumn{2}{|c||}{ \textbf{{Angular instability}} }
		&\multicolumn{4}{c||}{ \textbf{{Suboptimal values obtained by analytical results}} }&
		\multicolumn{4}{c| }{ \textbf{{Optimal values obtained by simulation results}} }\\
		\hline 
		%%------------------------------------------------------------- 
		%%-------------------------------------------------------------
		%%------------------------------------------------------------- Title
		$~~~\sigma_{to}~~~$&$\sigma_{ro}$&
		$~~N_{t}~~$&$~~N_{r}~~$&$\mathbb{P}_\textrm{out}$& Running time&
		$~~N_{t}~~$&$~~N_{r}~~$&$\mathbb{P}_\textrm{out}$& Running time \\
		\hline \hline 
		%%------------------------------------------------------------- 
		%%-------------------------------------------------------------
		%%------------------------------------------------------------- 1
		$5^o$&  $3^o$& 
		4&7&$1.52 \times 10^{-3}$&  $\simeq 1$ s&
		4&8&$1.35 \times 10^{-3}$& $\simeq 500$ s \\
		\hline
		%%------------------------------------------------------------- 
		%%-------------------------------------------------------------
		%%------------------------------------------------------------- 2
		$2^o$&  $4^o$& 
		9&5&$1.03 \times 10^{-4}$&  $\simeq 1$ s&
		9&5&$1.06 \times 10^{-4}$&  $\simeq 500$ s\\
		\hline
		%%------------------------------------------------------------- 
		%%-------------------------------------------------------------
		%%------------------------------------------------------------- 3
		$3^o$& $2^o$& 
		6&9&$2.85\times 10^{-5}$&  $\simeq 1$ s&
		6&9&$2.76 \times 10^{-5}$&  $\simeq 500$ s \\
		\hline
		%%------------------------------------------------------------- 
		%%-------------------------------------------------------------
		%%------------------------------------------------------------- 4
		$1^o$& $2^o$& 
		15&8&$2.39\times 10^{-7}$&  $\simeq 1$ s&
		\multicolumn{4}{c| }{$\mathbb{P}_\textrm{out}<10^{-6}$ and requires more independent runs } \\
		\hline
	\end{tabular}
	%}
	\label{tab1}%
\end{table*}%
%%%%%%%%%%%%%%%%%%%%%%%%%%%%%%%%%%%%
%%%%%%%%%%%%%%%%%%%%%%%%%%%%%%%%%%%
%
%

%
%
%%---- APPENDIX ---------------------------------------------------------------------
%%---- APPENDIX ---------------------------------------------------------------------
%%---- APPENDIX ---------------------------------------------------------------------
%
%
%%---- APPENDIX ---------------------------------------------------------------------
%%---- APPENDIX ---------------------------------------------------------------------
%%---- APPENDIX ---------------------------------------------------------------------
%
%
%%---- APPENDIX ---------------------------------------------------------------------
%%---- APPENDIX ---------------------------------------------------------------------
%%---- APPENDIX ---------------------------------------------------------------------

\appendices

%%-------------------------------------------------
%%-------------------------------------------------
\section{Proof of Theorem 1}
\label{AppA}
\begin{figure}
	\begin{center}
		\includegraphics[width=3.3 in ]{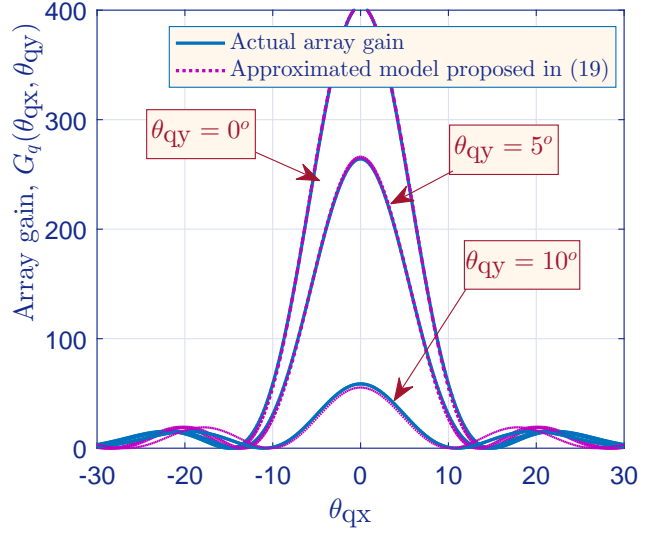}
		\caption{Approximated antenna pattern proposed in \eqref{xs} and comparing its validity with actual antenna pattern characterized in Section II for $N_q=8$.}
		\label{st4}
	\end{center}
\end{figure}

Note that $\gamma_{\textrm{uu}}$ is a function of five RVs $\alpha$, $\theta_{{tx}}$, $\theta_{{ty}}$, $\theta_{{rx}}$ and $\theta_{{ry}}$.	
To derive a tractable analytical model for $\gamma_{\textrm{uu}}$, we must first calculate the analytical model for 
$\mathbb{G}_{\textrm{uu}}(\theta_{{tx}},\theta_{{ty}},\theta_{{rx}},\theta_{{ry}})$
which is a functions of $ G_t(\theta_{{tx}},\theta_{{ty}}) $ and $G_r(\theta_{{rx}},\theta_{{ry}})$. As we observe from \eqref{p_1}, \eqref{vb1} and \eqref{f_1}, the array antenna gain $G_{{q}}$ is a complex function of $\theta_{qx}$ and $\theta_{qy}$, where subscript ${q}\in\{{t},{r}\}$ denotes, respectively, the Tx and Rx antenna. 
%-----------
After an exhaustive search over the actual pattern model provided in \eqref{p_1}, we obtain a simpler mathematical function for $G_\textrm{q}$ as
%\begin{align}
%\label{xs}
%G(\theta_x,\theta_y) \simeq 
%0.6366\times 	G_0(N)\,
% \frac{1-\cos( N k d_a \theta_{\textrm{q,xy}})}
%{N^2 \theta_{\textrm{q,xy}}^2},
%\end{align}
%where 
%\begin{align}
%\label{fg}
%\theta_{\textrm{q,xy}} = \sqrt{\theta_x^2 + \theta_y^2}.
%\end{align} 
\begin{align}
\label{xs}
G_{q}(\theta_{{qx}},\theta_{qy}) \simeq 
G_0''(N_{q})\,
\frac{1-\cos( N_{q} k d_a \sqrt{\theta_{qx}^2 + \theta_{qy}^2})}
{N_{q}^2 \left(\theta_{qx}^2 + \theta_{qy}^2\right)},
\end{align}
where $G_0''(N_{q})=0.2025\times 10^{\frac{G_{\textrm{max}}}{10}}G_0(N_{q})$.
For comparison with the actual antenna pattern obtained by \eqref{p_1}, the 3D graphical pattern generated by \eqref{xs} is plotted in Fig. \ref{x3}. For a better comparison, we also plot the 2D pattern generated by \eqref{xs} in Fig. \ref{st4} versus $\theta_{qx}$ for different values of $\theta_{qy}$. As we observe, an exact match exists between approximated model and actual antenna pattern, specially, at the main-lobe. 

Let us denote $\theta_{{q,xy}}= \sqrt{\theta_{qx}^2 + \theta_{qy}^2} $.
%
%As mentioned, for a general case, we consider a nonzero boresight error in addition to the random orientation fluctuations and model $\theta_\textrm{qx}$ and $\theta_\textrm{qy}$ as nonzero mean Gaussian distributed RVs, i.e., $\theta_\textrm{qx}\sim \mathcal{N}(\theta'_\textrm{qx},\sigma^2_\textrm{qo})$, and $\theta_\textrm{qy}\sim \mathcal{N}(\theta'_\textrm{qy},\sigma^2_\textrm{qo})$.
%
As mentioned, the RVs $\theta_{qx}$ and $\theta_{qy}$ are modeled as $\theta_{qx}\sim \mathcal{N}(\theta'_{qx},\sigma^2_{qo})$, and $\theta_{qy}\sim \mathcal{N}(\theta'_{qy},\sigma^2_{qo})$.
Hence, the PDF of $\theta_{{q,xy}}$ becomes Rician 
\begin{align}
\label{fg1}
f_{\theta_{{q,xy}}}(\theta_{{q,xy}}) \!=\! 
\frac{\theta_{{q,xy}}}{\sigma_{qo}^2}
\exp\!\left(\!-\frac{\theta_{{q,xy}}^2+\theta_{{q,xy}}'^2}{2\sigma_{qo}^2} \!\right)
I_0\!\left(\!\frac{\theta_{{q,xy}}\theta_{{q,xy}}'}{\sigma_{qo}^2}\!\right),
\end{align}
where $\theta_{{q,xy}}' = \sqrt{\theta_{qx}'^2+\theta_{qy}'^2}$, and $I_0(.)$ is the modified Bessel function of the first kind with order zero.
%----------
Moreover, Fig. \ref{p_1} clearly shows that the main power is radiated at main-lobe. Therefore, for a point-to-point link, it is reasonable to approximate the actual antenna array gain by only its main-lobe and for much more precision, main-lobe along with the first side-lobe. Now, by sectorizing \eqref{xs}, we propose a simpler sectorized model given by 
\begin{align}
\label{kgg}
G_{q}(\theta_{qx},\theta_{qy}) &\simeq G_{q}(\theta_{{q,xy}},D)= 
2 k^2 d_a^2 G_0''(N_{q})\,\Pi\left( D N_{q} \theta_{{q,xy}}  \right) \nonumber \\
%%%
+&G_0''(N_{q}) \sum_{i=1}^{j\,D-1}
\frac{D^2\left(1-\cos\left( \frac{i k d_a}{D}\right)\right)} {i^2}  \nonumber\\
%%%
\times& \left[ \Pi\left( \frac{D N_{q} |\theta_{{q,xy}}|}{i+1}  \right)  -  \Pi\left( \frac{D N_{q} |\theta_{{q,xy}}|}{i}  \right)  \right], 
\end{align}
%where $ i\in\{0,1,...,j(M-1)\}$ and $j\in\{1,2\}$ that $j=1$ is used when the main lobe of pattern is considered and for more precision, $j=2$ is used when main lobe along with the first side lobe is considered. 
where
$
\Pi(x)= \left\{
\begin{array}{rl}
1& ~~~ {\rm for}~~~ |x|\leq 1 \\
0& ~~~ {\rm for}~~~ |x|> 1 \\
\end{array} \right. 
$
and $j\in\{1,2\}$ whereby $j=1$ is used when the main-lobe of pattern is considered and for more precision, $j=2$ is used when main-lobe along with the first side-lobe is considered. 
For $j=1$, Figs. \ref{xx2}  and \ref{xx3} show the sectorized model for $D=6$ and $D=15$, respectively. Obviously, the accuracy of the proposed model increases by increasing $D$ at the cost of more complexity. Hence, choosing an optimal value for $D$ involves a tradeoff between tolerable complexity and desirable accuracy. Moreover, in Fig. \ref{xx4}, we show an example of sectorized model for $j=2$ and $D=10$.

%
%%%%%%%%%%%%%%%%%%%%%%%%%%%%%%%%%%%%%%%%%%%%%%%%%%%%%%%%%%%%%%%%
%%%%%%%%%%%%%%%%%%%%%%%%%%%%%%%%%%%%%%%%%%%%%%%%%%%%%%%%%%%%%%%% VERSUS P_T
%%%%%%%%%%%%%%%%%%%%%%%%%%%%%%%%%%%%%%%%%%%%%%%%%%%%%%%%%%%%%%%%
%%%%%%%%%%%%%%%%%%%%%%%%%%%%%%%%%%%%%%%%%%%%%%%%%%%%%%%%%%%%%%%%
%
%
%

From \eqref{fg1}, \eqref{kgg} and using \cite{math_wolfram}, after some mathematical manipulations, the PDF of $G_{q}(\theta_{qx},\theta_{qy})$ can be approximated as
\begin{align}
\label{op}
&f_{G_{q}}(G_{q}) =   
J_0(\theta'_{q,xy},\sigma^2_{qo})\,\delta\left(G_{q}-2 k^2 d_a^2 G_0''(N_{q}) \right)\\
%---------------------------
&~~~~~~~~~~~~~+\sum_{d_{q}=1}^{jD-1} J_{d_{q}}(\theta'_{q,xy},\sigma^2_{qo})\nonumber\\
%--------------------------- 
&~~~~~~~~~~~~~\times\delta\left(G_{q}- 
G_0''(N_{q}) \frac{D^2\left(1-\cos\left( \frac{d_{q} k d_a}{D}\right)\right)} {d_{q}^2}\right), \nonumber
\end{align}
where for $d_{q}\in\{0,1,...,jD-1\}$, we have
\begin{align}
\label{op1}
& J_{d_{q}}(\theta'_{q,xy},\sigma^2_{qo}) =
M\!\left(\frac{\theta'_{q,xy}}{\sigma_{qo}},\frac{d_{q}}{D N_{q} \sigma_{qo}}\right)
\!-\! M\left(\!\frac{\theta'_{q,xy}}{\sigma_{qo}},\frac{d_{q}+1}{D N_{q} \sigma_{qo}}\!\right)\!,
\end{align}
and $M(a,b)$ is the Marcum {\it Q}-function and can be formulated as
\begin{align}
\label{op2}
M(a,b) = \int_b^\infty x \exp\left(-\frac{x^2+a^2}{2} \right) I_0(ax).
\end{align}
Note that the Marcum {\it Q}-function is an standard function which can be readily computed.
From \eqref{pk1} and \eqref{op}, the PDF of RV $\mathbb{G}_{\textrm{uu}}(\theta_{{tx}},\theta_{{ty}},\theta_{{rx}},\theta_{{ry}})$ conditioned on RV $G_r(\theta_{{tx}},\theta_{{ty}})$ is derived as
\begin{align}
\label{pw1}
&f_{\mathbb{G}_{\textrm{uu}}|G_r} (\mathbb{G}_{\textrm{uu}}) = 
\frac{J_0(\theta'_{t,xy},\sigma^2_{to})}{G_r}
\,\delta\left(\frac{\mathbb{G}_{\textrm{uu}}}{G_r}
-2 k^2 d_a^2 G_0''(N_t) \right) + \nonumber\\
%---------------------------
&\sum_{d_{t}=1}^{jD-1} \!\frac{J_{d_{t}}(\theta'_{t,xy},\sigma^2_{to})}{G_r}
%--------------------------- 
\delta\!\left(\!\frac{\mathbb{G}_{\textrm{uu}}}{G_r}- 
G_0''(N_t) \frac{D^2\!\left(\!1\!-\!\cos\left( \frac{d_{t} k d_a}{D}\right)\!\right)} {d_{t}^2}\!\right). 
\end{align}
Using \eqref{op} and \eqref{pw1} and after some derivations, the PDF of $ \mathbb{G}_{\textrm{uu}}$ as a function of $N_t$, $N_r$, $\theta'_{t,xy}$, $\theta'_{r,xy}$, $\sigma^2_{to}$ and $\sigma^2_{to}$, is derived in \eqref{ro}. 
%
%
%%%%%%%%%%%%%%%%%%%%%%%%%%%%%%%%%%
%%%%%%%%%%%%%%%%%%%%%%%%%%%%%%%%%%
%\begin{figure*}[!t]
%	\normalsize
%	\begin{align}
%	\label{ro}
%	f_{\mathbb{G}_{\textrm{uu}}}(\mathbb{G}_{\textrm{uu}})&=
%	\mathbb{J}_{0,0}(\theta'_{t,xy},\theta'_{r,xy},\sigma^2_{to},\sigma^2_{ro})\, 
%	\delta\left(\mathbb{G}_{\textrm{uu}}
%	-4 k^4 d_a^4 G_0''(N_t) G_0''(N_r) \right)  \nonumber\\
%	%---------------------------
%	&~~~+\sum_{d_{t}=1}^{jD-1} 
%	\mathbb{J}_{d_{t},0}(\theta'_{t,xy},\theta'_{r,xy},\sigma^2_{to},\sigma^2_{ro})\,
%	%--------------------------- 
%	\delta\left(\mathbb{G}_{\textrm{uu}}- 
%	2 k^2 d_a^2 G_0''(N_t)
%	G_0''(N_r) \frac{D^2\left(1-\cos\left( \frac{d_{t} k d_a}{D}\right)\right)} {d_{t}^2}\right)  \nonumber \\
%	%---------------------------
%	%---------------------------
%	&~~~+\sum_{d_{r}=1}^{jD-1}  
%	\mathbb{J}_{0,d_{r}}(\theta'_{t,xy},\theta'_{r,xy},\sigma^2_{to},\sigma^2_{ro}) \, 
%	\delta\left(\mathbb{G}_{\textrm{uu}}
%	-2 k^2 d_a^2 G_0''(N_t) 
%	G_0''(N_{r}) \frac{D^2\left(1-\cos\left( \frac{d_{r} k d_a}{D}\right)\right)} {d_{r}^2}\right) \nonumber\\
%	%---------------------------
%	&~~~+\sum_{d_{r}=1}^{jD-1} \sum_{d_{t}=1}^{jD-1} 
%	\mathbb{J}_{d_{t},d_{r}}(\theta'_{t,xy},\theta'_{r,xy},\sigma^2_{to},\sigma^2_{ro}) \,
%	%--------------------------- 
%	\delta\left(\mathbb{G}_{\textrm{uu}}- 
%	4G_0''(N_t) G_0''(N_{r}) 
%	\frac{D^4\left(\sin^2\left( \frac{d_{t} k d_a}{2 D}\right)   \sin^2\left( \frac{d_{r} k d_a}{2 D}\right)   \right)} 
%	{d_{t}^2\,d_{r}^2}
%	\right).
%	\end{align}
%	\hrulefill
%	\vspace*{4pt}
%\end{figure*}
%%%%%%%%%%%%%%%%%%%%%%%%%%%%%%%%%
%%%%%%%%%%%%%%%%%%%%%%%%%%%%%%%%%
\begin{figure*}[!t]
	\normalsize
\begin{align}
\label{ro}
&f_{\mathbb{G}_{\textrm{uu}}}(\mathbb{G}_{\textrm{uu}})=
\mathbb{J}_{0,0}(\theta'_{t,xy},\theta'_{r,xy},\sigma^2_{to},\sigma^2_{ro})\, 
\delta\left(\mathbb{G}_{\textrm{uu}}
-4 k^4 d_a^4 G_0''(N_t) G_0''(N_r) \right)  \nonumber\\
%---------------------------
&+\sum_{d_{t}=1}^{jD-1} 
\mathbb{J}_{d_{t},0}(\theta'_{t,xy},\theta'_{r,xy},\sigma^2_{to},\sigma^2_{ro})\,
%--------------------------- 
\delta\left(\mathbb{G}_{\textrm{uu}}- 
2 k^2 d_a^2 G_0''(N_t)
G_0''(N_r) \frac{D^2\left(1-\cos\left( \frac{d_{t} k d_a}{D}\right)\right)} {d_{t}^2}\right)  \nonumber \\
%---------------------------
%---------------------------
&+\sum_{d_{r}=1}^{jD-1}  
\mathbb{J}_{0,d_{r}}(\theta'_{t,xy},\theta'_{r,xy},\sigma^2_{to},\sigma^2_{ro}) \, 
\delta\left(\mathbb{G}_{\textrm{uu}}
-2 k^2 d_a^2 G_0''(N_t) 
G_0''(N_{r}) \frac{D^2\left(1-\cos\left( \frac{d_{r} k d_a}{D}\right)\right)} {d_{r}^2}\right) \nonumber\\
%---------------------------
&+\sum_{d_{r}=1}^{jD-1} \sum_{d_{t}=1}^{jD-1} 
\mathbb{J}_{d_{t},d_{r}}(\theta'_{t,xy},\theta'_{r,xy},\sigma^2_{to},\sigma^2_{ro}) \,
%--------------------------- 
\delta\left(\mathbb{G}_{\textrm{uu}}- 
4G_0''(N_t) G_0''(N_{r}) 
\frac{D^4\left(\sin^2\left( \frac{d_{t} k d_a}{2 D}\right)   \sin^2\left( \frac{d_{r} k d_a}{2 D}\right)   \right)} 
{d_{t}^2\,d_{r}^2}
\right).
\end{align}
\hrulefill
\vspace*{4pt}
\end{figure*}
In \eqref{ro}, the parameter $\mathbb{J}_{d_{t},d_{r}}(\theta'_{t,xy},\theta'_{r,xy},\sigma^2_{to},\sigma^2_{ro})$ is defined as
\begin{align}
\label{ro1}
\mathbb{J}_{d_{t},d_{r}}(\theta'_{t,xy},\theta'_{r,xy},\sigma^2_{to},\sigma^2_{ro}) 
= J_{d_{t}}(\theta'_{t,xy},\sigma^2_{to}) J_{d_{r}}(\theta'_{r,xy},\sigma^2_{ro}).
\end{align}
Finally, from \eqref{ss1}, \eqref{Gamma} and \eqref{ro} and after some mathematical derivations, the PDF of RV $\gamma_{\textrm{uu}}$ is obtained in \eqref{uu}.
%%-------------------------------------------------
%%-------------------------------------------------

%Please refer to Appendix B.

%%%%%%%%%%%%%%%%%%%%%%%%%%%%%%%%%%%%%%%%%%%%%%%%%%%%%%%%%%%%%%
%%%%%%%%%%%%%%%%%%%%%%%%%%%%%%%%%%%%%%%%%%%%%%%%%%%%%%%%%%%%%%
%\bibliographystyle{IEEEtran}
%\balance
%\bibliography{IEEEabrv,myref}


\balance

\end{document}